# Room temperature Si:S barrier infrared detector with broadband response up to 4.4μm

He Zhu[1,#,*], Yunlong Xiao[1,2,#], Zhongyang Yu[1,#], Jiaqi Zhu[1], Qing Li[1,*], Zhenyu Ye[3], Xi Wang[1], Changlong Liu[1], Changyu Pan[1], Yufeng Shan[1], Junli Duan[1], Huizhen Wu[4], Weida Hu[1,2,*], Ning Dai[1,2,*]

**Abstract:** Mid-infrared spectrum is a critical tool for chemical analysis, industrial inspection, environment, and other fields due to its rich chemical bond information. However, the complicated growth or fabrication procedures of existing mid-infrared sensitive materials hinder the large-scale production and utilization of mid-infrared detectors. To address this issue, we developed Si:S barrier detectors employing sulfur doped silicon and a sophisticated band barrier design. Since the transport of dark current and photo current is separated, the barrier design effectively suppresses the dark current while allowing the photo current to leverage gain mechanisms, thereby substantially improving signal-to-noise ratio. As a result, the detector exhibits an infrared response range covering from 1.12 to 4.4μm with a peak at 3.3μm, excluding its intrinsic response in visible range. Its peak quantum efficiency surpasses that of the best mid-infrared silicon-based detector reported to date by an order of magnitude, reaching 2% at room temperature. The peak detectivity at 90K is $1.4\times10^{11}$ Jones @1.4V and decreases to $4.4\times10^{9}$ Jones @1.4V, 210K, comparable to the typical III-V and IV-VI photodetectors at one thousandth fabrication cost. Leveraging the well-established silicon-based manufacturing process, this device holds promise for large-scale production at a reduced price, offering a cost-effective solution for future mid-infrared detection.

## Introduction

Developing low-cost mid-infrared photodetector is of great importance for environment, chemical analysis, and industrial inspection [1-3]. Currently, mid-infrared detectors can be primarily classified into two categories: photon detectors and thermal detectors [4]. Photonic detectors, employing materials like mercury cadmium telluride (MCT) and indium antimonide (InSb), are the state-of-the-art mid-infrared detectors. But the growth of these materials is challenging, resulting in low yield and high cost [5-6]. After the 1980s, quantum well infrared detectors (QWIP) and type II superlattice (T2SL) detectors were developed using multivariate III-V compounds. However, QWIP have low quantum efficiency and demanding operating temperatures (~40K), while the production of T2SL material is limited by the substrate size [7-9]. Low-dimensional materials, such as quantum dots and 2D materials, have shown promising performance [10-14]. Nevertheless, they are in the early stage of laboratory research, with a considerable gap hindering large-scale production. Thermal detectors based on different kinds of photothermal effect offer economical alternatives but suffer from inferior detectivity compared to photodetectors [15]. Therefore, these technologies have yet to achieve cost-performance-balanced solutions for civilian applications.

With rich manufacturing experience and extensive manufacturing equipment, silicon-based detectors have the advantages of low cost and monolithic integration. However, the response of the silicon material is basically cut off at 1,100 nm subject to the bandgap limitation ($Eg$=1.12eV). Great efforts have been

1 Hangzhou Institute for Advanced Study, University of Chinese Academy of Sciences, Hangzhou, China. 2 State Key Laboratory of Infrared Physics, Shanghai Institute of Technical Physics, Chinese Academy of Sciences, Shanghai, China. 3 China Electronics Technology Group Corporation 38th Research Institute. 4 Institute of Physics, Zhejiang University, Hangzhou, China. e-mail: hezhu@ucas.ac.cn

invested in the heterostructure integration of IR-sensitive materials (including HgCdTe, InAsSb, PbSe, etc.) to take advantage of silicon manufacturing [16-20]. But heterostructure growth on silicon sacrifices CMOS compatibility and adds complexity to manufacturing. Extrinsic silicon, in which defects are introduced by ion implantation to enhance the absorption coefficient beyond the bandgap, offers another avenue for mid-infrared detection [21-22]. Since the thickness of the active region prepared in this way is limited, past attempts have primarily focused on increasing the doping concentration to improve its absorption coefficient. Consequently, most of the progress to date have leveraged hyperdoping silicon ($>10^{19}$cm$^{-3}$)[23-25]. Unfortunately, these photodetectors suffer from high dark current due to irreparable damage induced by the doping process [26-27]. To suppress the dark current, PN junction structure is introduced at the cost of the gain, which limits the quantum efficiency [26,28-29]. So far, no suitable method has obtained high quantum efficiency and low dark current simultaneously [30-31]. Additionally, despite adopting various elements such as Au, Zn, Te, the photoresponse of silicon-based detectors could barely be extended beyond 2.5μm, which is the cutoff wavelength for classical short-wavelength sensitive material [26,29,32-34].

In this study, we report a sulfur-doped Si barrier detector with room-temperature mid-infrared photoresponse covering from 1.12 to 4.4 μm. Fourier transform infrared spectrometer (FTIR) measurement demonstrates photoresponse peaks at 3.3μm, which is attributable to typical sulfur level 0.37eV below the conduction band minimum (CBM). Room temperature peak response of 44mA/W is achieved, corresponding to a quantum efficiency of 2%. The response time at room temperature is 30 μs, superior to thermal detectors. The dark current at temperature above 120K is effectively suppressed by a sophisticated band barrier design. Since the transportation of dark current and photocurrent is separated, the photocurrent can achieve gain while the dark current cannot. As a result, the peak detectivity $D^*_{\lambda_p}$ at 210K (achievable through thermoelectric (TE)-cooling) is $4.4\times10^9$ @1.4V and reaches $1.4\times10^{11}$ Jones at 90K, comparable to HgCdTe, InSb or type-II superlattice (T2SL) photodetector, etc [35]. Crucially, it is fabricated with CMOS-compatible technology, rendering it a competitive candidate for large-scale production in the future.

**Material characterization**

Fig. 1(a) illustrates the process of material fabrication involving ion implantation and rapid thermal annealing (RTA). This procedure entails multiple ion implantation steps with varying doses and energies to achieve uniform ion distribution. Detailed information regarding ion implantation and RTA are described in the device fabrication section. The annealed samples, prepared for high resolution transmission electron microscopy (HRTEM, Talos F200X G2) analysis, are fabricated by focused ion beam (FIB, Helios 5 UX) techniques as shown in Fig. S1(a). As shown by HRTEM result in Fig. 1(b), the crystal structure of the host silicon after RTA was effectively restored, exhibiting no discernible defects. The dopant profile after annealing was measured using secondary ion mass spectrometry (SIMS, TOF. SIMS Model 5-100). Fig. 1(c) reveals a flat impurity profile spanning from 20 nm to 0.8μm in-depth as expected, featuring an average concentration of approximately $2\times10^{17}$ cm$^{-3}$. Limited by the measurement accuracy of SIMS equipment, the data reliability is compromised from the surface to the depth of 20nm. Theoretical ionized carrier density and the Hall result is depicted in Fig. S2(a). The calculated carrier density is $1.9\times10^{15}$ cm$^{-3}$ at room temperature, consistent with the Hall result. This indicates most sulfur atoms are in substitution position and activated effectively. The comprehensive band structure derived from ab initio theoretical calculations is displayed in Fig. S3. As shown in the picture, silicon has a bandgap of 1.12eV, which restricts its absorption to wavelengths above 1100nm.

Due to the degeneracy effect of dopant atoms, an impurity band forms in the bandgap, denoted by green lines within the silicon bandgap. Photons with energies below 1.12 eV can be absorbed by the outer electrons of dopant atoms, thereby augmenting silicon's infrared absorption capability. The enhanced absorption effect is conspicuously evident in the FTIR absorption data presented in Fig. 1(e), where silicon's absorption extends significantly up to 7 μm.

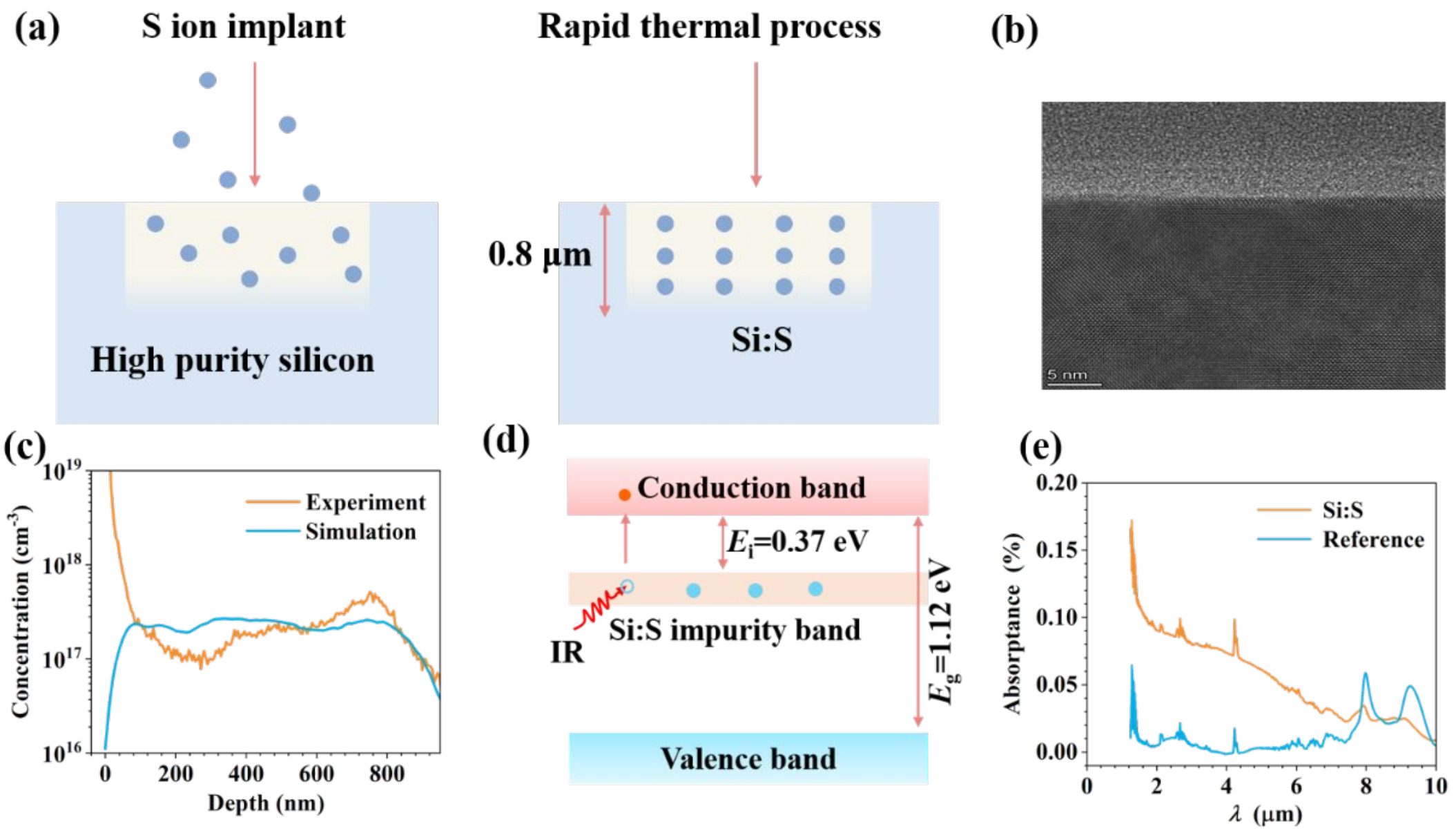

Figure 1 (a) The schematic of the implantation and annealing process (b) TEM result of Si:S sample after annealing (c) SIMS profile of Si:S sample after annealing (d) Simplified band structure of Si:S sample (e) Absorption measurement of annealed Si:S sample and reference silicon

**Electrical property of device**

The implanted sample are further fabricated into Si:S barrier detector with a sequence of nanofabrication procedures described in the device fabrication section. An optical microscope image of the fabricated sample is shown in Fig. 2(a). The detector comprises multiple basic units interconnected in parallel via interdigitated electrodes. As shown in in Fig. 2(b), each basic unit structure encompasses a sulfur-doped active region (AR), an intrinsic barrier region (BR), and two contact regions (CR). The active region and the barrier region extend over length of 450μm, with width of 20 μm and 5 μm, respectively. The sample prepared by FIB in Fig. S1(a) was further observed by TEM as shown in Fig. S1(b). Active region and barrier region can be clearly distinguished due to difference in the thickness of the covered silicon oxide on those areas. The diffraction pattern and an enlarged image of the area enclosed by the white box in Fig. S1(b) are presented in Fig. S1(c)&(d), respectively. These results demonstrate the excellent crystal quality of active region. Subsequently, the electrical properties of the sample were comprehensively investigated. As shown in Fig. 2(c), electrons transmit from the dopant atoms to the unintentional compensated acceptors, leaving behind empty levels in the impurity band. Since the distance between two dopant atoms is close enough, the electron can hop from one atom to the empty level of another atom without being activated into the conduction band. The motion of electrons within the impurity band can be considered as the movement of holes in the opposite direction. Therefore, those empty levels within the impurity band are termed "quasi-holes" due to their hole-like behavior. The transport of quasi-holes in the impurity band forms hopping current that are not attenuated by cooling [36]. As a result, silicon photoconductors subjected to heavy doping or hyper-doping exhibit high dark currents, even being cooled to critically low temperatures. While classical pn junction structure is commonly employed to suppress hopping current, its disadvantage lies in the cost of the photoconductive gain, which further have a negative influence on quantum efficiency. Therefore, we introduce a band barrier structure rather than pn junction structure to reconcile the conflict between dark current and quantum efficiency. An intrinsic region is inserted between the active region and the contact region to cut off the impurity energy band. To evaluate the electrical property, the dark current $I_d$ of the barrier detector is measured with a semiconductor parameter analyzer (Keithley 6517B). The dark current at each temperature is positively correlated with bias. The temperature-dependent dark current of the device at 0.1 V is extracted and drawn in Fig. 2(d) to investigate the dark current component. Notably, the dominant components of dark current vary with temperature. When the temperature is below 150K, hopping current is the predominant dark current component. To further elucidate the effect of the barrier, the simulated I-V curves of a reference photoconductor at different temperature is displayed in Fig. S4(a) and S4(b). At temperatures below 120K, the predominant dark currents of the barrier detector and the reference device are both hopping currents. Once the temperature is elevated above 120K, the dominant dark current of the photoconductor becomes the generation-recombination current due to the increased thermally activated carriers in the conduction band. The behavior differs for the barrier detector. The depletion of quasi-hole in the active region induced by the disconnectivity of the impurity band suppresses the recombination effect. The dominant current remains as a hopping current until 160K for the barrier detector. The depletion region gradually fails due to enhanced thermal excitation as the temperature surpasses 160 K [37]. Subsequently, the generation-recombination current becomes the dominant component [38, 39]. This approach effectively suppresses the dark current compared to a photoconductor structure.

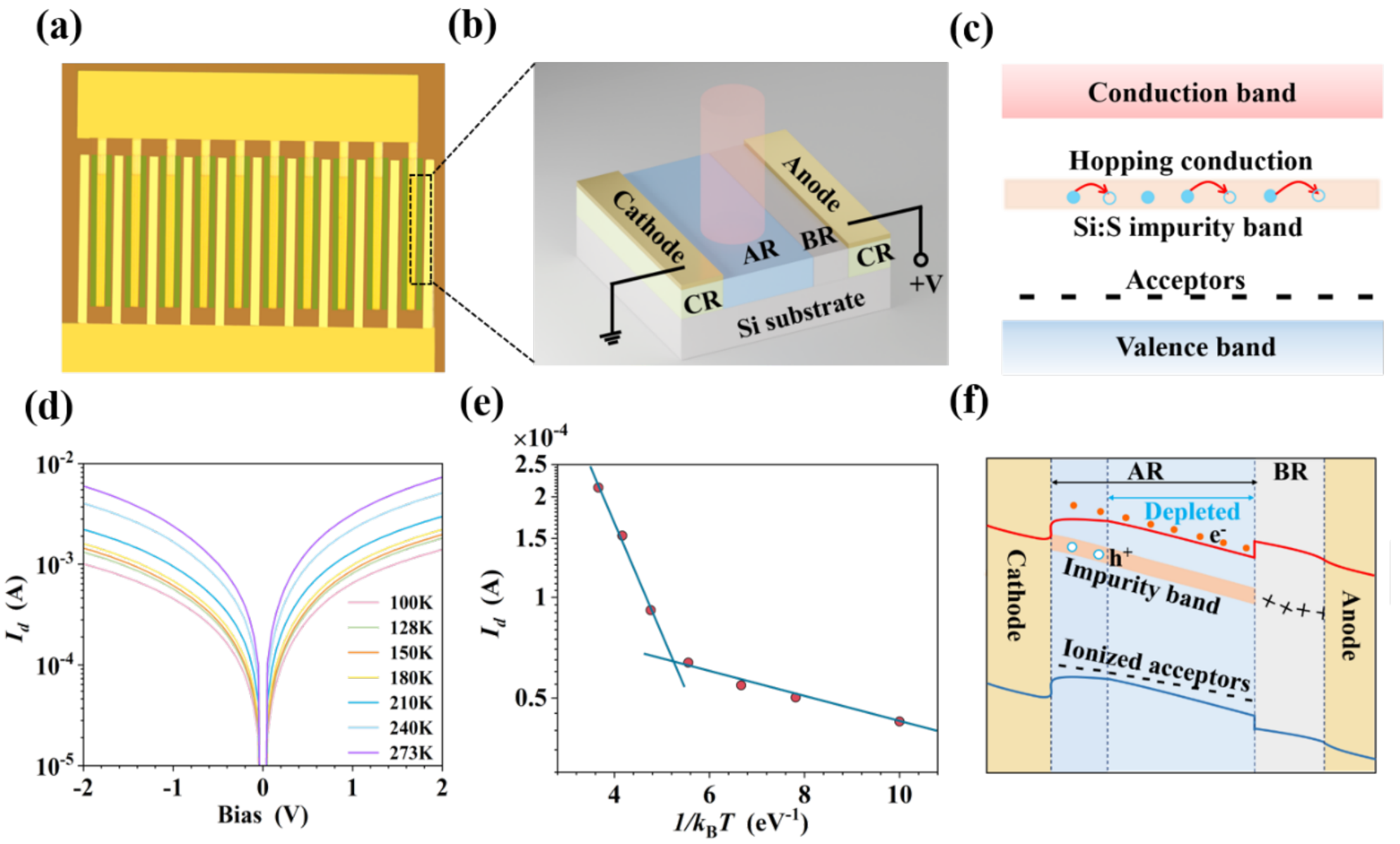

Figure 2 (a) Optical image of the fabricated Si:S barrier detector (b) Schematic of the one unit in the interdigital electrode detector (c) Hopping current mechanism of Si:S barrier detector (d) Experimental I-V of Si:S barrier detector under different temperatures (e) Temperature-dependent dark current of Si:S barrier detector under 0.1V (f) Band structure of Si:S barrier detector under operation mode

**Optoelectronic property of device**

The operational mechanism under illumination is illustrated in Fig. 3(a). When the detector works under light illumination, photo-generated carriers will be pumped into the conduction band in the active region. These carriers, transported in the conduction band, can leverage two different gain mechanisms, including the multiplication effect owing to impact ionization [40] and the photoconductive gain attributed to the injection carriers from the electrode. Conversely, the dark current associated with quasi-holes is confined to the impurity band, precluding the utilization of these mechanisms. As a result, the signal-to-noise ratio, denoted as detectivity, is improved. The responsivity $R_{bb}$ in Fig. 3(b) is assessed with a 1000K blackbody source (RCN1250). The specific measurement setup is outlined in blackbody response section and Fig. S5. As anticipated, the responsivity increases with bias at each temperature point. Below 150K, it slightly increases with temperature, likely attributed to thermally assisted activation. The responsivity decreases rapidly with rising temperature above 150K. Consequently, the maximum blackbody responsivity appears at 150K with a value of 1.13A/W @2V. The spectral response of the device was measured using FTIR (Bruker 80v) equipment, as illustrated in Fig. S6. As shown in Fig. 3(c), sulfur doping successfully extends the response spectra of the silicon barrier detector beyond the intrinsic bandgap (~1100 nm), reaching ~4.4 μm. The peak response is observed at 3.3 μm, corresponding to the classical sulfur level 0.37eV below CBM [21]. The spectral response intensity exhibits an increasing trend with temperature up to 150K, followed by a rapid decrease above 150K, consistent with the variation observed in blackbody responsivity. Due to the insufficient intensity of the source in the FTIR equipment, the response at room temperature is measured using a tunable pulsed mid-infrared laser. The FTIR spectral response denoted by the orange line in Fig. 3(d) was obtained at 150K, with the intensity scaled down to facilitate comparison with the results at room temperature. The responsivity profile at room temperature aligns well with the response spectrum measured by FTIR, with intensity at least an order of magnitude higher than the best-known room-temperature mid-infrared silicon detector, as shown in Fig. 3(d) [22]. The external quantum efficiency (EQE) is calculated by the formula $\eta = \hbar cR/\lambda q$, in which $h$ represents Planck constant, $c$ represents velocity of light, $q$ represents unit charge. The result in Fig. 3(d) indicates the room temperature EQE of 2% at 3.3μm.

The response-frequency relationship in Fig. 3(e) and fall time at varying-temperatures in Fig. 3(f) were assessed utilizing a 3.85μm quantum cascade laser (QCL). The laser light undergoes modulation by a laser controller (Thorlabs ITC4005QCL), generating a square wave with a constant duty cycle of 50%. The rise and fall time of the modulated laser are both ~2μs. The experimental configuration employed for measuring response-frequency relationship closely resembled the setup in Fig. S5, with the light source being replaced. The 3dB bandwidth at room temperature is 8.8 kHz, corresponding to a response time of 40 μs. The bandwidth at 80K is decreased to 0.47kHz, attributing to the prolonged carrier lifetime. On the contrary, nanosecond-scale carrier lifetime in hyperdoping samples [27, 41] implies the existence of numerous recombination defects. The rise and fall time were measured with a similar setup, with the lock-in amplifier being substituted by an oscillator (4455F, 2G bandwidth). As shown in Fig. S7(c), the rise time is shorter than the fall time. Therefore, the response time, which serves an indicator of carriers' lifetime, depends on the fall time. The fall time at 80K is 745μs, consistent with the calculation from the 3dB bandwidth. It fluctuates around 700μs below 160K as shown in Fig. 3(f). Thereafter, it rapidly declined to 31μs at room temperature as depicted in Fig. S7(d). The decrease in lifetime is conceivably related to an augmented thermal ionization donor concentration, a trend which aligns relatively consistently with the response. To estimate the gain value, the cross-time of the photo-generated carrier in Fig. 3(f) is calculated with the mobility data from Fig. S2(b) [42, 43]. Further details are provided in the gain calculation section. Given that mobility experiences only minor variations within this temperature range, the cross time of the photo-generated carrier is maintained at ten nanoseconds. With the response time and carrier's cross-time, the photoconductive gain could be calculated as presented. Additionally, the second component of gain related to the multiplication effect is also calculated by the formula in the gain calculation section. The temperature-dependent gain is plotted in Fig. 3(g). Given the response is proportional to the gain, the trend of the response is consistent with the gain. Although the adopted doping concentration is two orders of magnitude lower than hyper-doping silicon, the gain of several thousand times yields a quantum efficiency one order higher than the hyperdoping silicon detector.

The blackbody detectivity is calculated by the formula $D_{bb}^{*} = R_{bb}\sqrt{A_d}/\sqrt{\frac{4k_BT\Delta f}{R} + 2qI_d\Delta f}$, in which $A_d$, $\Delta f$, $k_B$, $T$, $R$ represents effective photosensitive area, bandwidth, Boltzmann constant, operating temperature, resistance, respectively. The former part $\sqrt{\frac{4k_BT\Delta f}{R}}$ is defined as thermal noise (or Johnson noise), while the latter $\sqrt{2qI_d\Delta f}$ is the shot noise, with $\Delta f$ typically set at 1Hz [44, 45]. The value $D_{bb}^{*}$ correlates positively with the operating bias at each temperature as shown in Fig. 3(h). Over the range of 90K-150K, temperature slightly influences the blackbody detectivity. Above 150K, the blackbody detectivity experienced a rapid decline. Specifically, at 90K, the blackbody detectivity reached approximately $6.3\times10^{9}$ Jones @1.4V. The spectral detectivity is calculated through $D_{\lambda}^{*} = D^{*} * g * R_{\lambda}/R_{\lambda_p}$, in which $g$, $R_{\lambda}$ indicates g factor and spectral responsivity, respectively. The calculation of $g$ and $R_{\lambda}/R_{\lambda_p}$ is discussed in spectral response section. The result is compared with the typical III-V and IV-VI mid-infrared detectors in Fig. 3(i). Notably, the peak detectivity at a temperature of 90K is $1.4\times10^{11}$ Jones @1.4V, close to the performance of HgCdTe and InSb photodetectors. Like most photodetectors, the performance degrades with increasing temperature. However, owing to an effective dark current suppression mechanism, our device still achieves a pretty good detectivity of $4.4\times10^{9}$ Jones @1.4V at 210K, achievable through thermoelectric cooling. In a word, the performance of our device is good enough to serve as a substitute in cost-sensitive scenarios.

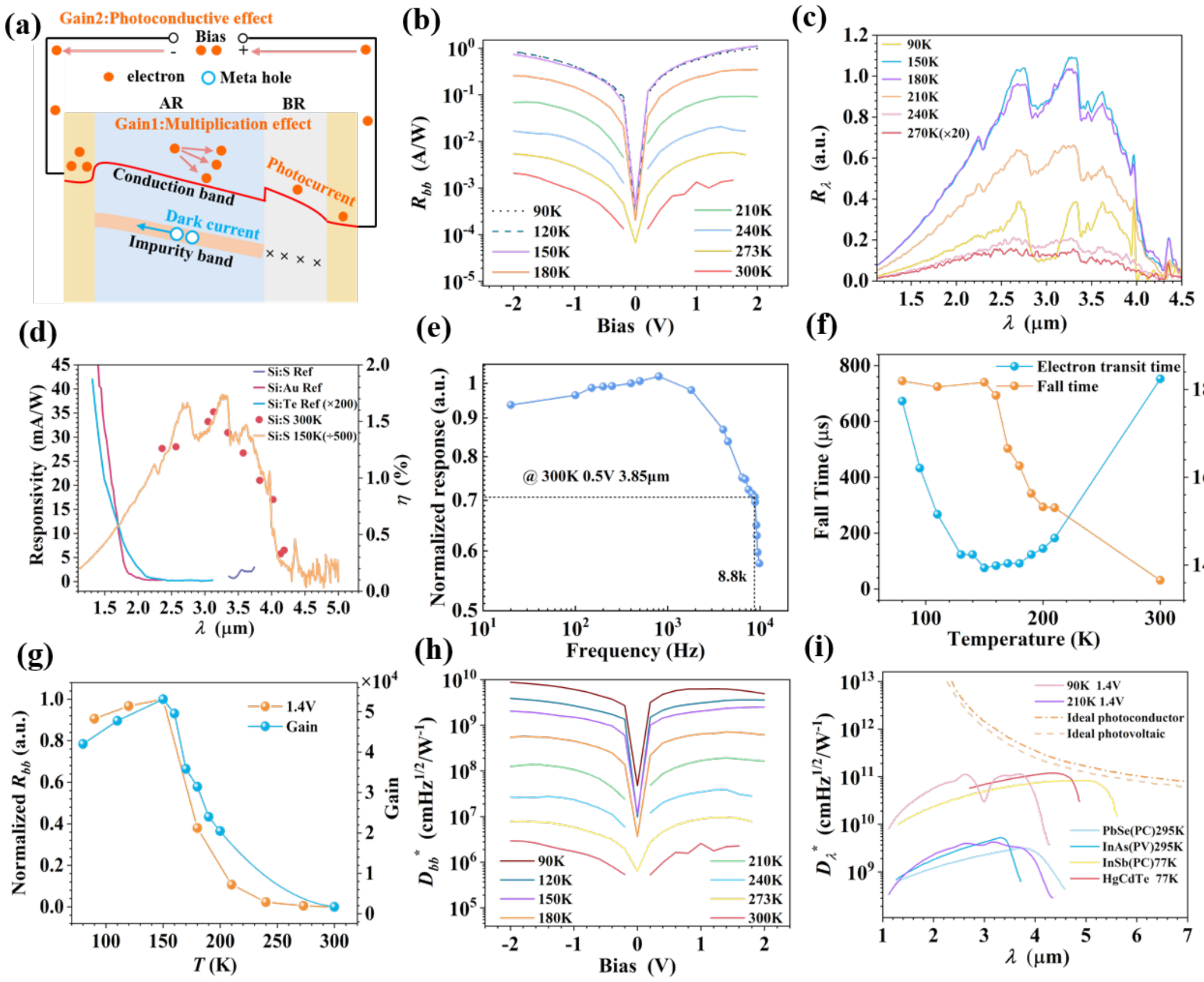

Figure 3 (a) Schematic drawing of two different gain mechanisms of photo-generated carrier (b) Bias-dependent blackbody responsivity of Si:S barrier detector under different temperature (c) Relative spectral response of Si:S barrier detector at 1.4V under different temperatures (d) Dot: Spectral responsivity and quantum efficiency measured with tunable laser at 1V under 300K,Orange Solid line: Normalized relative spectral response at 1.4V under 210K (e) 3dB bandwidth of Si:S barrier detector measured at 1.4V under 300K (f) Fall time and cross time of Si:S barrier detector measured at 1.4V under different temperatures (g) Blackbody responsivity and photoconductive gain versus temperature at 1.4V (h) Bias-dependent blackbody detectivity of Si:S barrier detector under different temperature (i) Comparison of the spectral detectivity with different detectors

## Photocurrent mapping and imaging performance

The results of the laser beam-induced photocurrent (LBIC) are presented in Fig. 4. Our experimental system comprises two distinct light sources: a tungsten lamp, serving as the reference light source for sample observation, and a replaceable laser utilized as the signal light source for photocurrent measurements. The lasers employed are 637nm, 1550nm, and 4000nm for Fig. 4(b), (c), and (d), respectively. All mapping tests are conducted at 80K. For ease of comparison, the data in Fig. 4(b)~(d) is extracted from the same square region with a side length of 80 μm delineated by the white dash line in Fig. S8(b). As evidenced by those images, the infrared response originates from the detector area. In the LBIC system, a refractive objective is utilized for laser wavelengths below 1800 nm, whereas a reflective objective is employed for wavelengths exceeding 1800 nm. The alteration in optical setup in Figure 4(d) results in a slight spatial shift in the mapping area. Nevertheless, the clear photocurrent mapping in Fig. 4(d) vividly demonstrates the exceptional mid-infrared response of the barrier detector. Since the laser spots in Fig. 4 have dimensions on the order of tens of microns, they are insufficient for discriminating between the barrier region and the active region. However, the interface between the barrier region and the anode become more and more sharp with the extension of the wavelength. The profile of photocurrent along the dotted line in Fig. S8(d)~(f) at different biases is depicted in Fig. 4(e). At low bias, the active region is incompletely depleted. The electrical field in non-depleted active region can be neglected. Therefore, the further the distance from the depletion region, the more challenging it is for the photo-generated carriers to be transported to the electrode, yielding a diminished photocurrent. With the elevation of the operation bias, the active region is gradually completely depleted, amplifying the multiplication effect for carriers generated further away from the AR/BR interface. As a result, regions exhibiting the strongest response gradually migrate away from the interface.

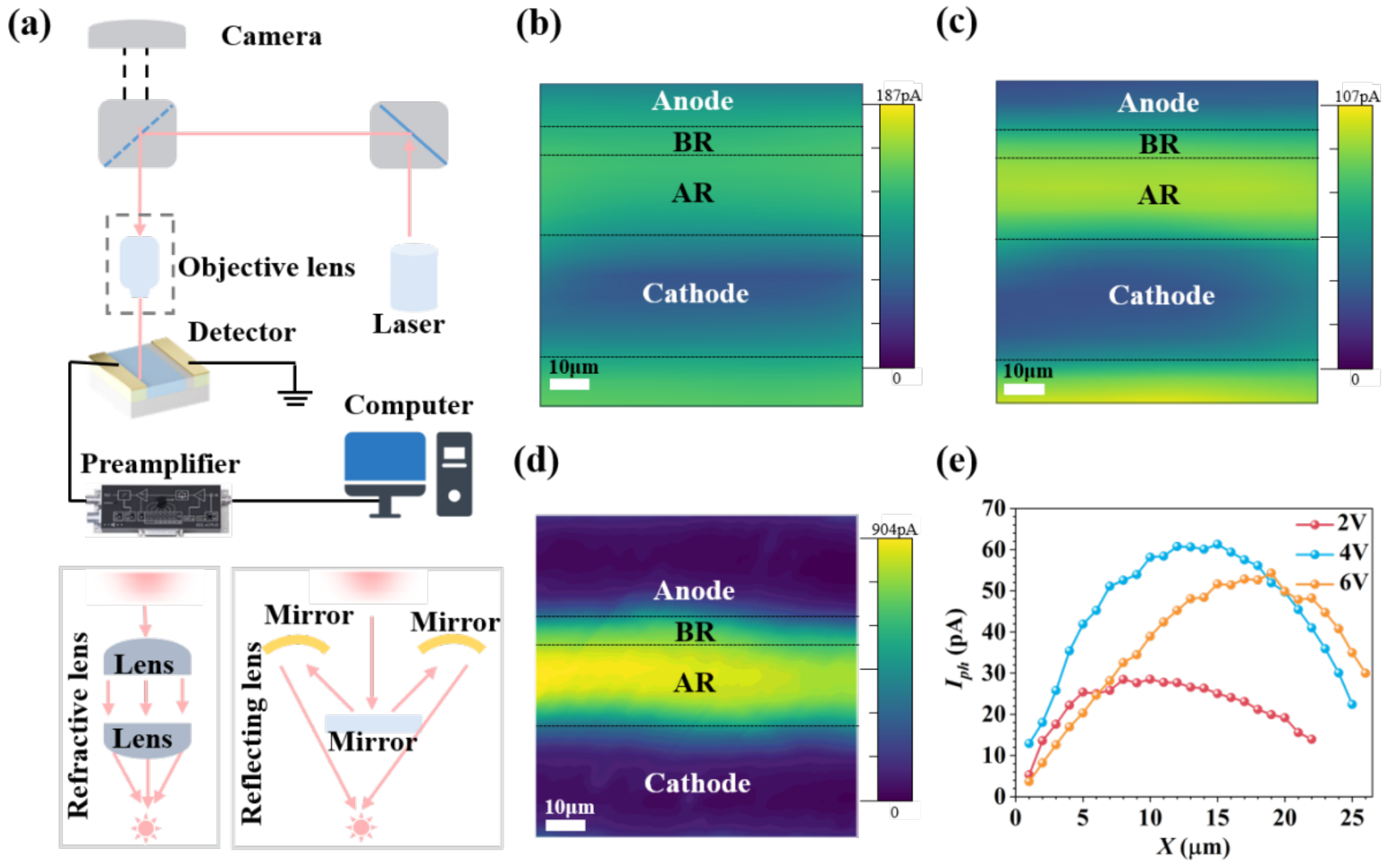

Figure 4 (a) Laser beam induced photocurrent (LBIC) setup (b)~(d) LBIC mapping result of Si:S barrier detector pumped by 637nm, 1550nm and 4000nm laser, respectively (e) LBIC profiles of Si:S barrier detector under different operation bias along the dashed line in Fig. S8

The imaging performance was evaluated with a scanning system, as illustrated in Fig. 5(a). The target object is mounted on a two-dimensional scanning stage. A $BaF_2$ lens is used to focuses the light from the target on the detector. The reference visible and scanning transmitted images of three cuvettes containing different liquid compositions are displayed in Fig. 5(b) and Fig. 5(c), respectively. The right cuvette is empty while the left and the middle cuvettes contained isopropyl alcohol (IPA) and tetrachlorethylene (PCRE), respectively. As liquids exhibit high transparency to visible light, all liquids appear transparent in Fig. 5(b). The 3.85μm QCL laser is employed as the light source for the measurement depicted in Fig. 5(c). As indicated by the FTIR spectra in Fig. S9, IPA exhibits strong absorption at 3.85 μm, while PCRE shows minimal absorption. Consequently, PCRE remains transparent in Fig. 5(c), whereas the cuvette containing IPA appears opaque. Additionally, imaging of an electric soldering was performed, with a temperature of approximately 470 K and an emission peak at 6.4 μm. The distinct thermal image of the electric soldering also confirms the superior mid-infrared performance of the Si:S barrier detector. However, the space resolution is limited by the resolution of the scanning stage and the device structure, resulting in the distortion of the electric soldering.

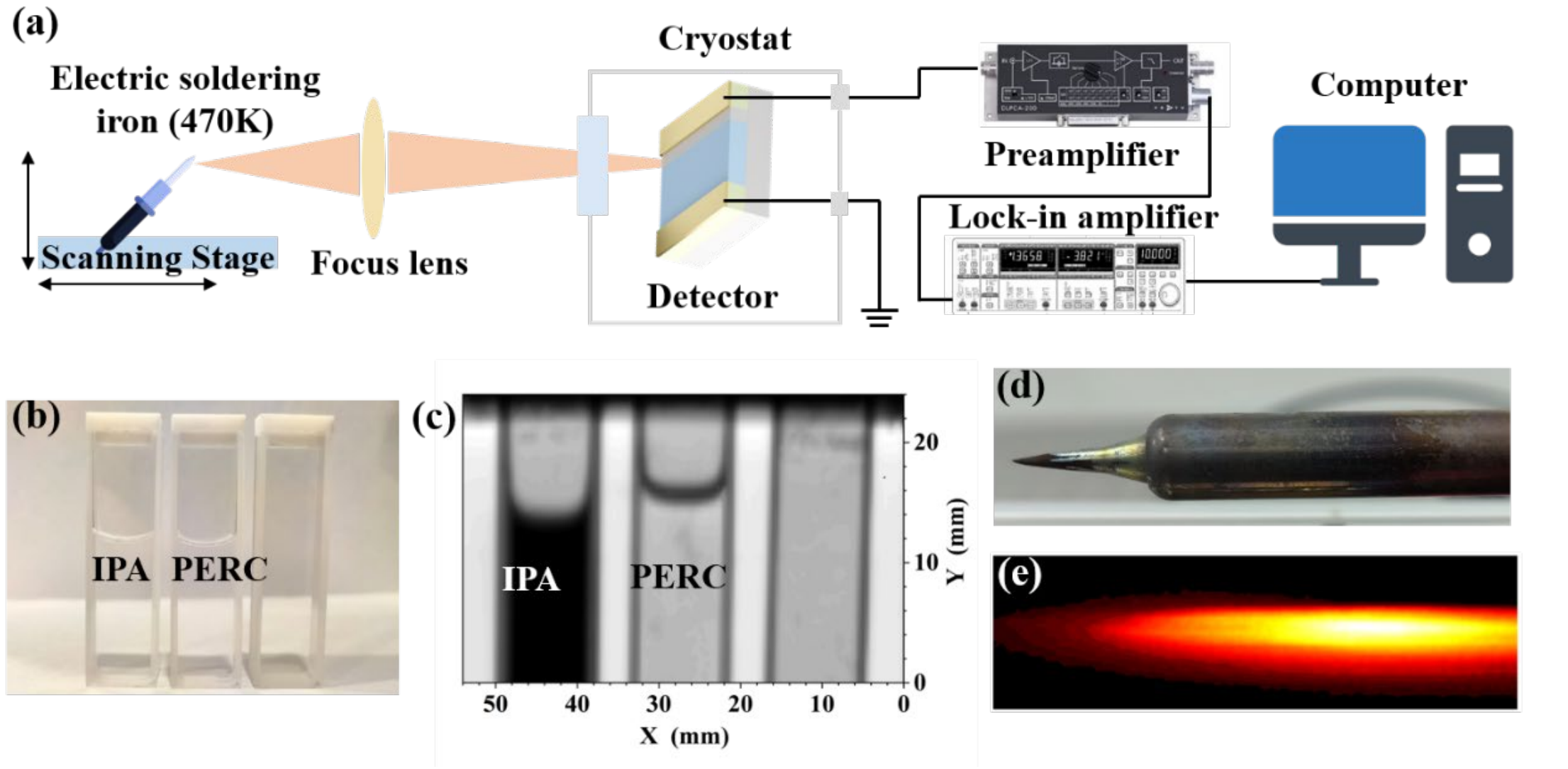

Figure 5 (a) Scanning image demonstration setup (b)~(c) Imaging demonstration of the barrier detector for three cuvettes containing different visible transparent chemicals, including IPA, PERC, and air (d)~(e) Imaging demonstration of the barrier detector for a 470K electric soldering iron

The excellent performance of Si:S barrier detector has been conclusively demonstrated in the aforementioned tests. Beyond performance, the practical utility of the detector hinges on various factors, including reliability and cost. Despite the commendable performance of emerging materials, including two-dimensional materials[46] and quantum dots[47], their applicability is hampered by poor reliability attributable to their active chemical properties. The reliability of our device is evaluated in Fig. S10. The device is stored in a standard ambient air for over five years without any special treatment. Spectral response assessments were conducted in the years 2018 and 2023, revealing minimal alterations in spectral profiles over the intervening years across the wavelength range of 1.12 to 4.4 μm. These findings collectively underscore the remarkable reliability of our device. Additionally, the operational life can be significantly extended through proper preservation methods, such as hermetic sealing within a vacuum transistor outline (TO) package. The information on the response range and manufacturing costs of commercial mid-infrared detectors are concluded in Figure S11a. Benefitting from mature semiconductor manufacturing technology, the cost of Si:S barrier detectors is at least two orders of magnitude lower than the existing mid-infrared detectors. Although the performance at room temperature necessitates improvement, the performance at thermoelectric cooling temperatures already rivals that of commercial detectors such as InAsSb. Simultaneously, ultra-large-scale arrays can be manufactured based on our detector structure by exploiting large-size high-quality silicon materials, thus providing unique advantages over other materials. As shown in Figure S11(b), beyond traditional military, scientific research, and industrial fields, we forecast that silicon-based mid-infrared detectors featuring lower prices and larger array scale will be expanded into more price-sensitive scenarios, such as transportation, exploration, smart home-in facilities, and mobile phones, etc.

This study also has limitations. From 2018 to the present, we have conducted a total of three experiments. However, only one experiment successfully fabricated samples exhibiting mid-infrared response at room temperature, yielding a success rate of 33%. Other sample batches solely responded to short-infrared light, as evidenced in Fig. S12(a). These samples exhibited sulfur aggregation as depicted in Figure S11(b). Analysis of the fabrication process revealed several potential factors influencing the success rate. Specifically, the ion implanter used failed to replicate previous implantation energy and may have been contaminated due to frequent disassembly. Furthermore, the annealing equipment procedure has been modified since it's shared equipment. However, as supported by reference [42], observing a broad-band mid-infrared response for Si:S samples at 90K is justifiable. Reference [22] substantiates the feasibility of room temperature mid-infrared response for Si:S samples. These studies uniformly employ the PN junction structure for designing mid-infrared photodetectors. While this design effectively suppresses dark current, it comes at the expense of gain, leading to either a poor responsivity or negative result at room temperature in the mid-infrared range. To strike a balance between dark current and responsivity, we introduced a barrier structure, resulting in an improved signal-to-noise ratio at room temperature. Despite encountering challenges with success probability, our findings align consistently with existing literature. Consequently, after careful deliberation, we have opted to publish our current results.

**Conclusion**

In this work, a room temperature mid-infrared Si:S barrier detector is developed through sulfur implantation and band barrier design. Remarkably, the performance of the barrier detector is outstanding among all the reported mid-infrared silicon detectors. Its photo-response covers 1.12 to 4.4μm. The response peak is located at 3.3μm, corresponding to the characteristic sulfur energy level positioned 0.37eV below CBM. Our device exhibits the peak quantum efficiency of around 2% at room temperature, a noteworthy enhancement compared to the reported mid-infrared silicon-based photodetectors. The peak detectivity at a temperature of 90K is $1.4\times10^{11}$ Jones @1.4V and decreases to $4.4\times10^{9}$ Jones @1.4V, 210K, comparable to the typical III-V and IV-VI photodetectors. The exceptional performance of our device can be primarily attributed to the innovative band design. At temperatures above 120K, the barrier could effectively suppress the dominant dark current component without sacrificing of photoconductive gain. By leveraging the separated transportation route, the photo-generated carrier could experience a giant gain while the dark carrier cannot. Furthermore, distinguishable imaging of three cuvettes containing different materials and a clear outline of electric soldering also demonstrates the excellent infrared performance of our device. In the future, taking advantage of mature silicon manufacturing technology, large-scale production of pixel detectors and focal plane arrays could be realized. In conclusion, our device emerges as a compelling contender among a diverse array of mid-infrared detectors.

**Methods**

**[Device Fabrication]**

Initially, a 100-orientation intrinsic silicon wafer with resistivity exceeding $10^4$ Ω·cm is coated with 200nm silicon oxide by Plasma Enhanced Chemical Vapor Deposition (PECVD, oxford 80+) equipment. The silicon oxide atop the active region is selectively removed with hydrofluoric acid, in preparation for the ion implantation. Five sequential implantation procedures, each varying in dopant dose and energy, are meticulously executed to ensure a uniform distribution of dopant atoms. Following the active region implantation, the contact region is formed using phosphorous as the dopant with a similar protocol. The implantation dose in the contact region is two magnitudes higher than the active region. Subsequent to the completion of the implantation process, the sample is subjected to RTA (RTP 150) at 950℃ for 30 seconds to active the dopant and repair the implantation damage. The silicon substrate between the active region and contact region plays the role of barrier region. Aluminum electrode with 600nm thickness is deposited on the contact region using a radio sputter (JSS-450-1). After being treated in vacuum at 450℃ for 30 minutes, ohmic contact is achieved. To assess the device's performance, gold wire bonding is established by Westbond 7476D.

**[Gain calculation]**

The gain of photogenerated carrier arises from two distinct mechanisms, including photoconductive gain $G_1$ and the multiplication effect gain $G_2$:

$$G = G_1 \cdot G_2$$

The expression of the first part of gain factor is:

$$G_1 = \tau/t_{cross} \approx \tau_{fall}/t_{cross}$$

Here, the cross time of the calculated by

$$t_{cross} = \frac{L}{v} = \frac{L}{\mu E}$$

in which $L, v, \mu, E$ represents the width of the detector, the velocity of the photo-generated carrier [42, 43], the mobility of the photo-generated carrier and the electrical field, respectively. The electrical field can be determined by the TCAD software as depicted in Fig. S4(d).

The expression of the second part of gain factor is [36]:

$$G_2 = \int_0^{\omega} g_{op}(x) M(x) dx \,/\, \int_0^{\omega} g_{op}(x) dx$$

in which $g_{op}(x), M(x)$ represents the generation rate of carriers and the multiplication factor, respectively.

The expression of carrier generation rate is given by [48]:

$$g_{op}(x) = \int \alpha \Phi (1 - \mathrm{R}_1) e^{-\alpha z} dz$$

where $\alpha, \Phi, R_1, z$ are absorption coefficient, incident luminous flux, surface reflectivity, implantation depth, respectively.

The expression of the multiplication factor is:

$$M(x) = e^{\int_0^x \xi(x') dx'}$$

$\xi(x)$ is the impact ionization coefficient for the electrons [49].

**Data availability**

The data that support the conclusions of this study are available from the corresponding authors upon reasonable request. Source data are provided with this paper.

**Code availability**

Any codes used in this study are available from the corresponding authors upon request.

**Reference**

1. Dyrek, Achrène. et al. SO2, silicate clouds, but no CH4 detected in a warm Neptune. *Nature* 625(7993), 51-54, (2024).
2. Liu, Zhu. et al. Near-real-time monitoring of global CO2 emissions reveals the effects of the COVID-19 pandemic. *Nature Communications* 11(1), 5172, (2020).
3. Kim, Hyungjin. et al. Actively variable-spectrum optoelectronics with black phosphorus. *Nature* 596(7871), 232-237, (2021).
4. Rogalski, A. Infrared detectors: status and trends. *Progress in Quantum Electronics* 27(2-3),

59-210, (2003).

5. Madejczyk, Paweł. et al. MCT heterostructures for higher operating temperature infrared detectors designed in Poland. *Opto-Electronics Review* 31, (2023).
6. Rogalski, A. History of infrared detectors. *Opto-Electronics Review* 20, 279-308, (2012).
7. Palaferri, Daniele. et al. Room-temperature nine-µm-wavelength photodetectors and GHz-frequency heterodyne receivers. *Nature* 556(7699), 85-88, (2018).
8. Maimon S. and G. W. Wicks. nBn detector, an infrared detector with reduced dark current and higher operating temperature. *Applied Physics Letters* 89.15, (2006).
9. Jia, C. et al. Antimonide-based high operating temperature infrared photodetectors and focal plane arrays: a review and outlook. *Journal of Physics D: Applied Physics* 56, (2023).
10. Rogalski, A. Detectivities of WS2/HfS2 heterojunctions. *Nature Nanotechnology* 17(3), 217-219, (2022).
11. Zhu, Jiaqi. et al. Lateral photovoltaic mid-infrared detector with a two-dimensional electron gas at the heterojunction interface. *Optica* 7(10), 1394-1401, (2020).
12. Koepfli, Stefan M. et al. Metamaterial graphene photodetector with bandwidth exceeding 500 gigahertz. *Science* 380.6650, 1169-1174, (2023).
13. Tang, X. et al. Dual-band infrared imaging using stacked colloidal quantum dot photodiodes. *Nature Photonics* 13, 277-282, (2019).
14. Chen, Y. et al. Unipolar barrier photodetectors based on van der Waals heterostructures. *Nature Electronics* 4, 357-363, (2021).
15. Tan, X. et al. Non-dispersive infrared multi-gas sensing via nanoantenna integrated narrowband detectors. *Nature communications* 11, 5245, (2020).
16. Lin, H. et al. Mid-infrared integrated photonics on silicon: a perspective. *Nanophotonics* **7**, 393-420, (2017).
17. De Lyon, T. J. et al. MBE growth of HgCdTe on silicon substrates for large-area infrared focal plane arrays: a review of recent progress. *Journal of electronic materials* 28, 705-711, (1999).
18. Delli, E. et al. Mid-Infrared InAs/InAsSb Superlattice nBn Photodetector Monolithically Integrated onto Silicon. *ACS Photonics* 6, 538-544, (2019).
19. Xu, H. et al. Imaging of the Mid-Infrared Focal Plane Array With a Modified Read-Out Circuit. *IEEE Transactions on Instrumentation and Measurement* 72, 1-6, (2023).
20. Goossens, Stijn. et al. Broadband image sensor array based on graphene–CMOS integration. *Nature Photonics* 11, 366-371, (2017).
21. Janzén, Eo. et al. High-resolution studies of sulfur-and selenium-related donor centers in silicon. *Physical Review B* 29, 1907-1918, (1984).

22. Fain, Romy. et al. CMOS-compatible mid-infrared silicon detector. CLEO: Science and Innovations Optica Publishing Group, (2017).
23. Umezu, I. et al. Emergence of very broad infrared absorption band by hyperdoping of silicon with chalcogens. *Journal of Applied Physics* 113. 21, (2013).
24. Sullivan, J. T. et al. Methodology for vetting heavily doped semiconductors for intermediate band photovoltaics: A case study in sulfur-hyperdoped silicon. *Journal of Applied Physics* 114, (2013).
25. Simmons, C. B. et al. Enhancing the Infrared Photoresponse of Silicon by Controlling the Fermi Level Location within an Impurity Band. *Advanced Functional Materials* 24, 2852-2858, (2014).
26. Wang, M. et al. Extended Infrared Photoresponse in Te-Hyperdoped Si at Room Temperature. *Physical Review Applied* 10, (2018).
27. Sher, M. J. et al. Picosecond carrier recombination dynamics in chalcogen-hyperdoped silicon. *Applied Physics Letters* 105, (2014).
28. Tabbal, Malek. et al. Formation of single crystal sulfur supersaturated silicon based junctions by pulsed laser melting. *Journal of Vacuum Science & Technology B: Microelectronics and Nanometer Structures Processing, Measurement, and Phenomena* 25.6, 1847-1852, (2007).
29. Mailoa, J. P. et al. Room-temperature sub-band gap optoelectronic response of hyperdoped silicon. *Nat Commun* 5, 3011, (2014).
30. Adinolfi, V. & Sargent, E. H. Photovoltage field-effect transistors. *Nature* 542, 324-327, (2017).
31. Saran, R. & Curry, R. J. Lead sulphide nanocrystal photodetector technologies. *Nature Photonics* 10, 81-92, (2016).
32. García-Hemme, E. et al. Room-temperature operation of a titanium supersaturated silicon-based infrared photodetector. *Applied Physics Letters* 104.21, (2014).
33. Ackert, J. J. et al. High-speed detection at two micrometres with monolithic silicon photodiodes. *Nature Photonics* 9, 393-396, (2015).
34. Berencen, Y. et al. Room-temperature short-wavelength infrared Si photodetector. *Scientific reports* 7.1, 43688, (2017).
35. Teledyne Judson Technologies https://www.teledynejudson.com/home.
36. Szmulowicz, F. & Madarasz, F. L. Blocked impurity band detectors—an analytical model: Figures of merit. *Journal of Applied Physics* 62, 2533-2540, (1987).
37. Zhu, H. et al. Temperature-sensitive mechanism for silicon blocked-impurity-band photodetectors. *Applied Physics Letters* 119, (2021).
38. Liao, K. S. et al. Extended mode in blocked impurity band detectors for terahertz radiation

detection. *Applied Physics Letters* 105, (2014).

39. Pan, C. et al. Dark-Current-Blocking Mechanism in BIB Far-Infrared Detectors by Interfacia l Barriers. *IEEE Transactions on Electron Devices* 68, 2804-2809, (2021).
40. Pan, C. et al. Observation of gain operation mode in Ge:B BIB THz detector. *AIP Advanc es* 11.5, (2021).
41. Streetman, B. G. Carrier Recombination and Trapping Effects in Transient Photoconductive Decay Measurements. *Journal of Applied Physics* 37, 3137-3144, (1966).
42. Migliorato P, Elliott C T. Sulphur doped silicon IR detectors[J]. *Solid-State Electronics*, 21. 2, 443-447, (1978).
43. Klaassen, D. B. M. A unified mobility model for device simulation—II. Temperature depen dence of carrier mobility and lifetime. *Solid-State Electronics* 35.7, 961-967, (1992).
44. Fang, Yanjun. et al. Accurate characterization of next-generation thin-film photodetectors. *Na ture Photonics* 13.1, 1-4, (2019).
45. Wang, Fang. et al. How to characterize figures of merit of two-dimensional photodetectors. *Nature Communications* 14.1, 2224, (2023).
46. Huang, Y. et al. Interaction of Black Phosphorus with Oxygen and Water. *Chemistry of Ma terials* 28, 8330-8339, (2016).
47. Gréboval, C. et al. Mercury Chalcogenide Quantum Dots: Material Perspective for Device I ntegration. *Chemical Reviews* 121, 3627-3700, (2021).
48. Zhu, H. et al. Optimized Si-Based Blocked Impurity Band Detector Under Alternative Oper ational Mode. *IEEE Transactions on Electron Devices* 66, 3891-3895, (2019).
49. Lotz, Wolfgang. Electron-impact ionization cross-sections and ionization rate coefficients for atoms and ions from hydrogen to calcium. *Zeitschrift für Physik* 216.3, 241-247, (1968).
50. Sze, S. M. et al. Physics of semiconductor devices. John wiley & sons, (2021). .
51. Zhu, Jiaqi. et al. Ultrabroadband and multiband infrared/terahertz photodetectors with high s ensitivity. *Photonics Research* 9.11, 2167-2175, (2021).
52. Zhao qiang. Introduction to the Basics of Infrared Sensors. http://share.hamamatsu.com.cn/sp ecialDetail/988.html.
53. Product Information-Optical Sensor https://www.hamamatsu.com.cn/cn/zh-cn/product/optical-sen sors.html.
54. Product Information https://vigophotonics.com/product/pc-2te-5/
55. Product Information https://vigophotonics.com/product/pv-2te-4/
56. Product Information https://www.hamamatsu.com/jp/en/product/optical-sensors/infrared-detector/i nsb-photovoltaic-detector/P5968-100.html
57. Product Information https://www.hamamatsu.com/jp/en/product/optical-sensors/infrared-detector/i

nas-photovoltaic-detector/P7163.html

58. Product Information https://www.hamamatsu.com/jp/en/product/optical-sensors/infrared-detector/inassb-photovoltaic-detector/P16612-011CN.html
59. Delli, E. et al. Mid-Infrared InAs/InAsSb Superlattice nBn Photodetector Monolithically Integrated onto Silicon. *ACS Photonics* 6, 538-544, (2019).
60. Product Information https://www.hamamatsu.com/jp/en/product/optical-sensors/infrared-detector/type2-superlattice/P15409-901.html
61. Liu, X. H. et al. Effects of bias and temperature on the intersubband absorption in very long wavelength GaAs/AlGaAs quantum well infrared photodetectors. *Journal of Applied Physics* 115, (2014).
62. Product Information https://www.nepcorp.com/PbSe_TEC.html
63. Peterson, J. C. & Guyot-Sionnest, P. Room-Temperature 15% Efficient Mid-Infrared HgTe Colloidal Quantum Dot Photodiodes. *ACS Applied Materials & Interfaces* 15, 19163-19169, (2023).
64. Chen, M., Hao, Q., Luo, Y. & Tang, X. Mid-Infrared Intraband Photodetector via High Carrier Mobility HgSe Colloidal Quantum Dots. *ACS Nano* 16, 11027-11035, (2022).
65. Bullock, J. et al. Polarization-resolved black phosphorus/molybdenum disulfide mid-wave infrared photodiodes with high detectivity at room temperature. *Nature Photonics* 12, 601-607, (2018).
66. Long, Mingsheng, et al. Room temperature high-detectivity mid-infrared photodetectors based on black arsenic phosphorus. *Science advances* 3.6 e1700589 (2017).
67. Peng, M. et al. Room-Temperature Blackbody-Sensitive and Fast Infrared Photodetectors Based on 2D Tellurium/Graphene Van der Waals Heterojunction. *ACS Photonics* 9, 1775-1782, (2022).
68. Liang, Q. et al. High-Performance, Room Temperature, Ultra-Broadband Photodetectors Based on Air-Stable PdSe2. Advanced Materials 31, e1807609, (2019).
69. Mailoa J P, Akey A J, Simmons C B, et al. Room-temperature sub-band gap optoelectronic response of hyperdoped silicon[J]. *Nature communications*, 5(1): 3011 (2014).
70. Fu J, Cong J, Cheng L, et al. Zinc-hyperdoped silicon photodetectors fabricated by femtosecond laser with sub-bandgap photoresponse. *Semiconductor Science and Technology*, 37(12): 124004 (2022).
71. Huang S, Cao J, Song G, et al. Broadband-Spectral-Responsivity of black silicon photodetector with high gain and sub-bandgap sensitivity by titanium hyperdoping. *Optics & Laser Technology*, 171: 110399 (2024).
72. Berencén Y, Prucnal S, Liu F, et al. Room-temperature short-wavelength infrared Si photod

etector. *Scientific reports*, 7(1): 43688 (2017).

73. Jia Z, Wu Q, Jin X, et al. Highly responsive tellurium-hyperdoped black silicon photodiode with single-crystalline and uniform surface microstructure. *Optics Express*, 28(4): 5239-5247 (2020).
74. Zhang K, He J, He T, et al. Extended infrared responses in Er/O-hyperdoped Si at room temperature. *Optics Letters*, 46(20): 5165-5168 (2021).
75. Fain, Romy, et al. CMOS-compatible mid-infrared silicon detector. CLEO: Science and Innovations. Optica Publishing Group, (2017).

**Acknowledgements**

This work was financially supported by National Natural Science Foundation of China (62175045, 11933006, U2141240); National Key Research and Development Project of China (2023YFB2806700); Hangzhou Science and Technology Bureau (TD2020002); National Key Laboratory Foundation (SKLIP2021006); Research Funds of Hangzhou Institute for Advanced Study (B02006C019019, 2022ZZ01007). We thanks Professor Huixiong Deng and Professor Junwei Luo from Institute of Semiconductors, Chinese Academy of Sciences for the discussion of the band structure, thanks Professor Zhiyong Tan from the Shanghai Institute of Microsystem and Information Technology, Chinese Academy of Sciences for the spectral response measurement.

**Competing interests**

The authors declare no competing interests.

## Supplementary Material

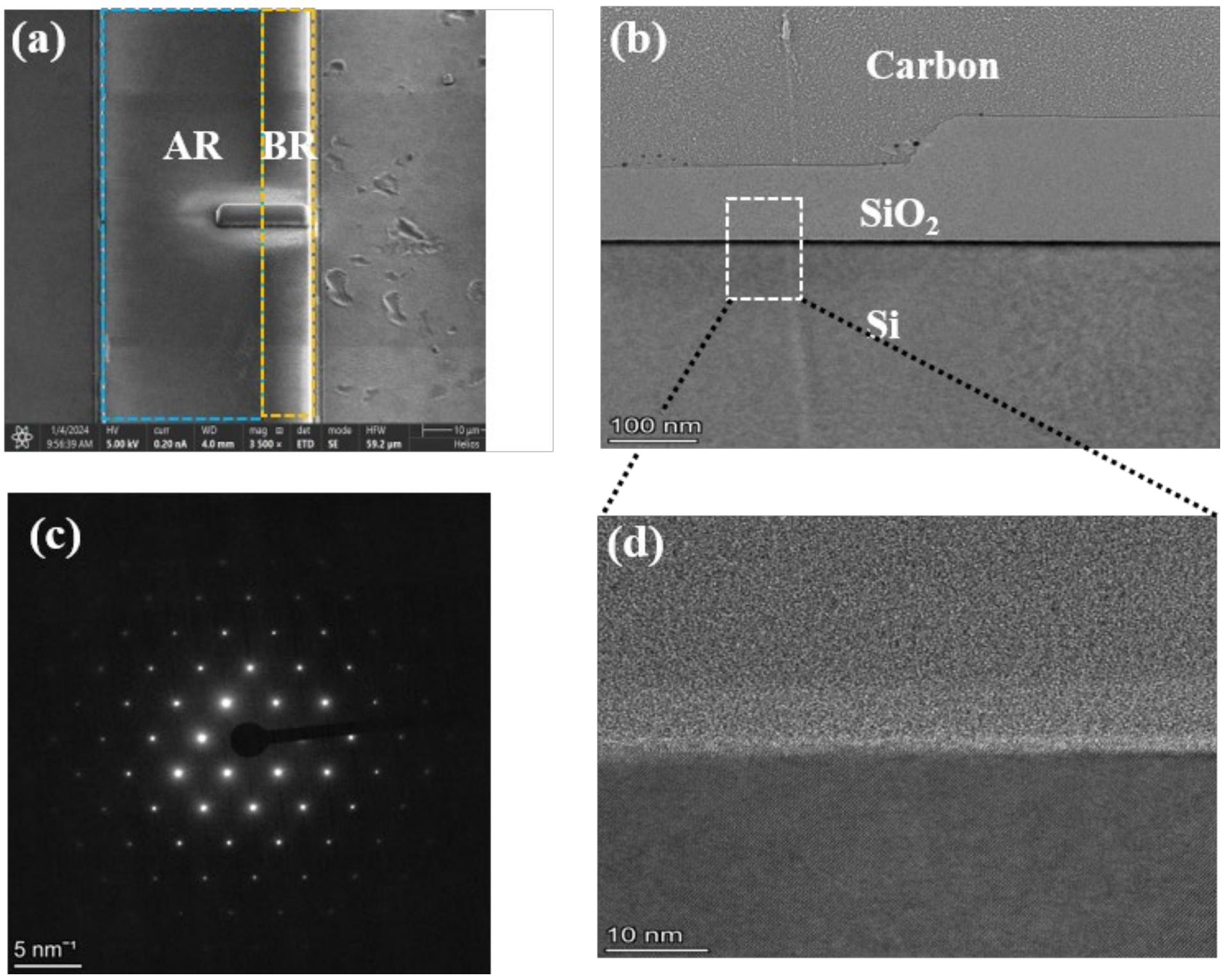

Fig. S1 (a) SEM image during the TEM sample preparation process (b,d) TEM pictures of cross-sections of sample in (a) with different magnification (c) Diffraction pattern obtained from the same region in (d)

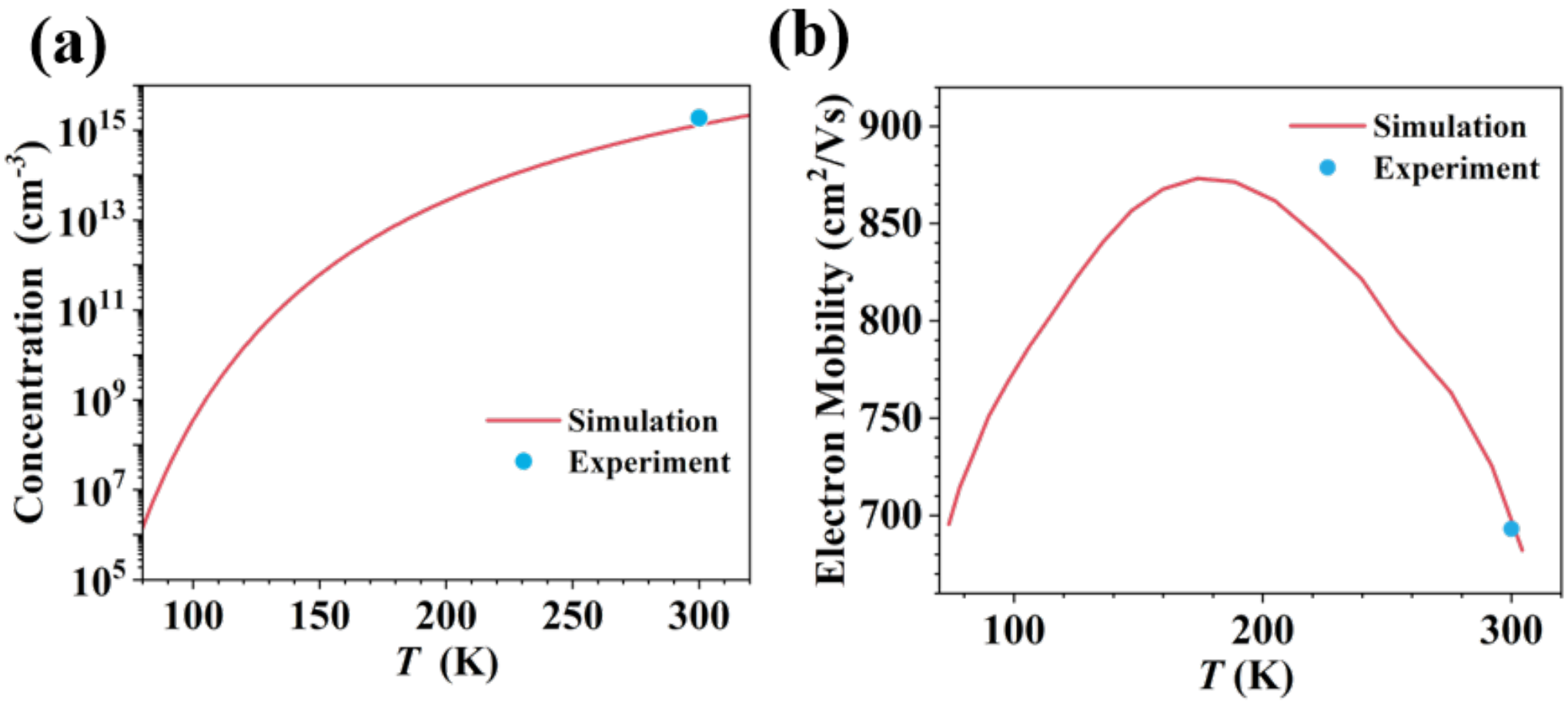

Fig. S2 (a) The calculated carrier density of annealing Si:S sample under different temperature and the Hall effect result (b) Calculated carrier mobility of annealing Si:S sample under different temperature and the Hall effect result

The carrier density is calculated with the formula $n = (\frac{N_D N_C}{2})^{1/2} \exp{(-\frac{\Delta E_D}{2k_0 T})}$ [50], in which $N_D, N_c, \Delta E_D, k_0, T$ represent acceptor concentration, effective density of states in conduction band, activation energy, Boltzmann constant and temperature, respectively. The mobility of the carrier is calculated by the Klaassen Model. [42,43]

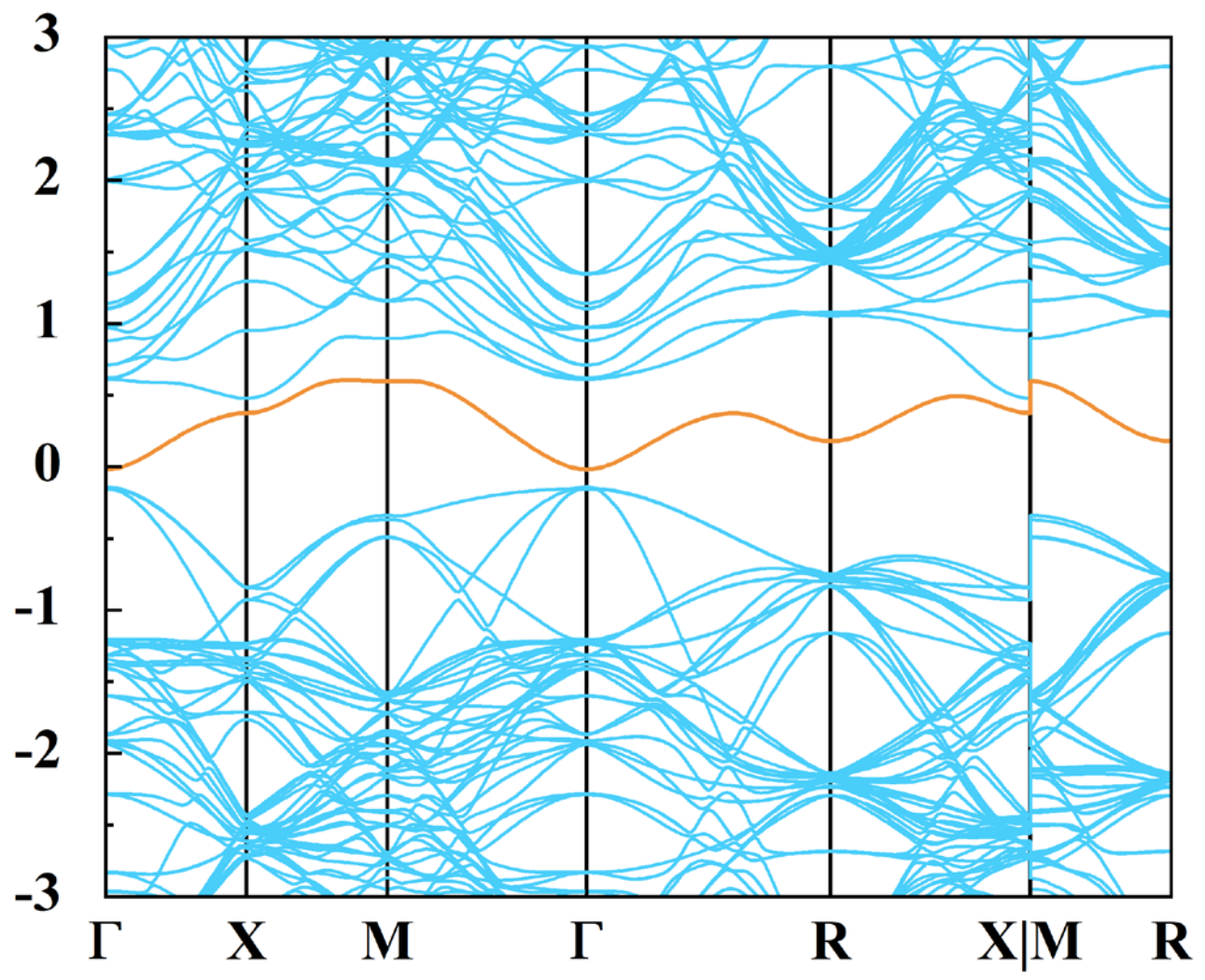

Fig. S3 Band structure of Si:S sample calculated by ab-initio theory, the green line represents the impurity band

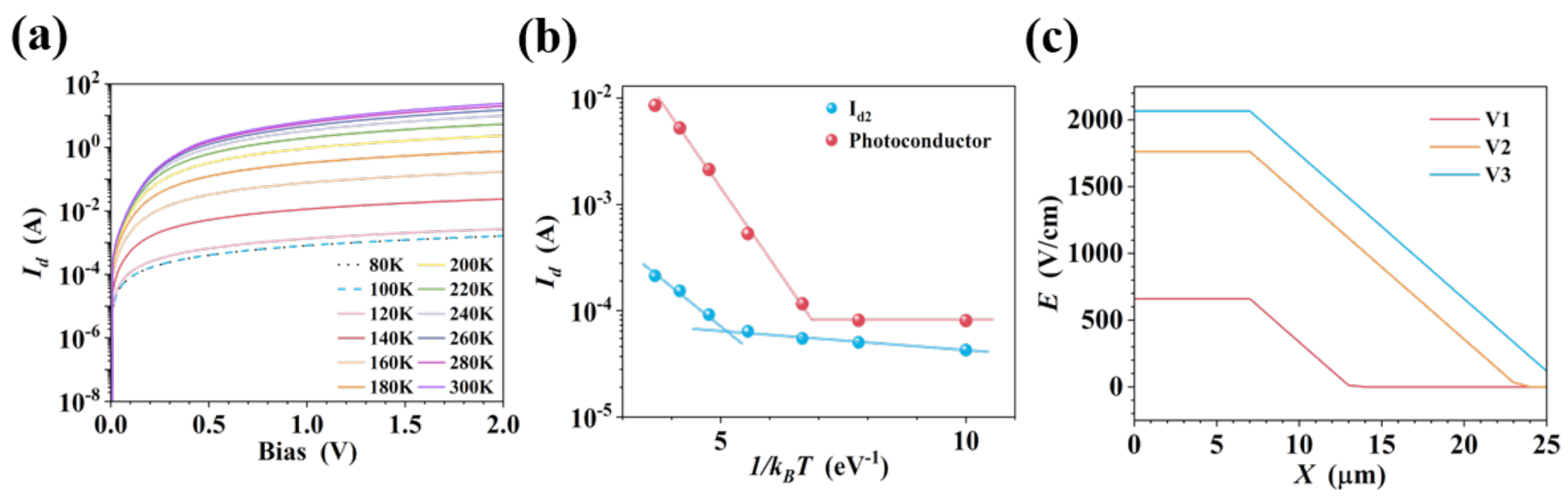

Fig. S4 (a) Simulated I-V curves of the reference photoconductive detector at different temperatures (b) Temperature-dependent dark current of Si:S barrier detector and reference photoconductive detector at 0.1V (c) The electrical field distribution in Si:S barrier detector under different bias, the origin of the coordinate locates at the anode/BR interface

The current in Fig. S4(a) and the electrical field in Fig. S4(c) are calculated using the Lumerical Device software. The structure of the Si:S barrier detector in Fig. S4(c) is the same as the tested detector, featuring a 5μm-width barrier region and a 20μm-width active region. The barrier region is replaced by the doped region in the reference photoconductive detector, thus resulting in a 25μm-width active region.

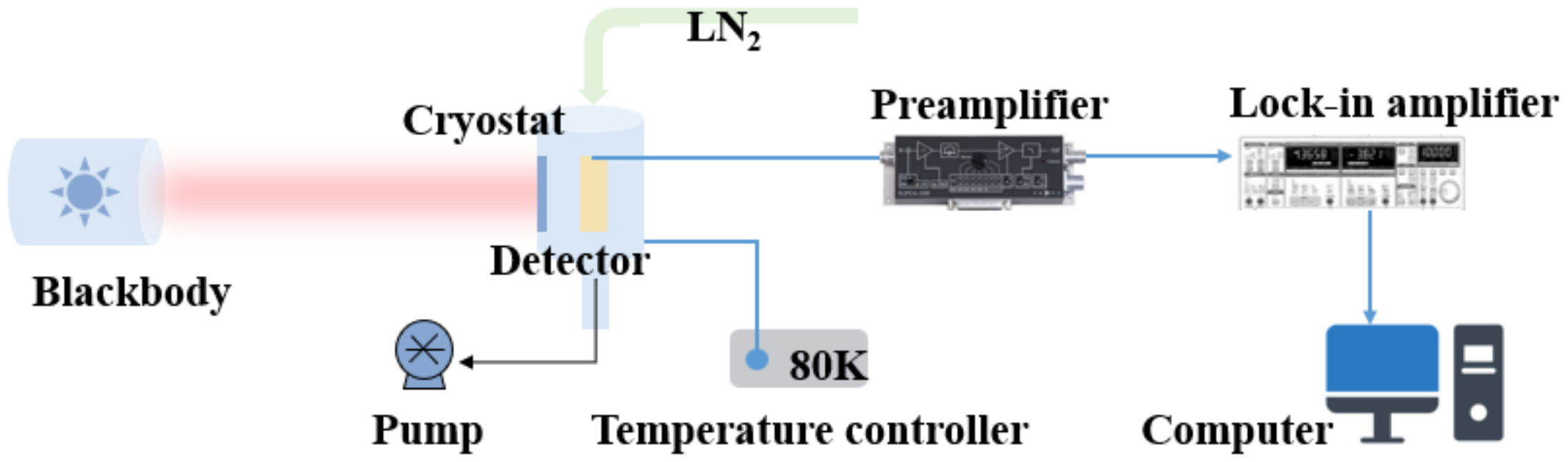

Fig. S5 Schematic of the measurement setup for blackbody response. The window material of the cryostat is KRS5, which is transparency in the range from 0.6 to 40μm

Blackbody response was measured using a calibrated commercial blackbody furnace (HFY-200B) in conjunction with a 1200nm long pass filter to avoid the intrinsic response of silicon [37]. The schematic of the measurement setup is plotted in Fig. S6. The emitted radiation is modulated into a 237Hz AC signal by a mechanical chopper (Stanford SR540). The device was placed in a dewar (Janis ST100) placed 12 cm away from the aperture. The temperature is controlled by Lakeshore Model 325. The photocurrent of the detector is amplified by a preamplifier (Stanford SR570) before input into a lock-in amplifier (Stanford SR830). The total incident power on the device surface can be calculated using the approximate formula $P = \varepsilon\alpha\sigma(T_b^4 - T_d^4)A_bA_d/\pi L^2$, where $\varepsilon$=0.99 is average emissivity, α is the modulation factor, $T_b$ is the blackbody temperature, $T_d$ is the background temperature, σ is the Stefan-Boltzmann constant, $A_b$ is the blackbody exit aperture area, $A_d$ is the effective photosensitive area, and $L$ is the distance from the detector to the blackbody.

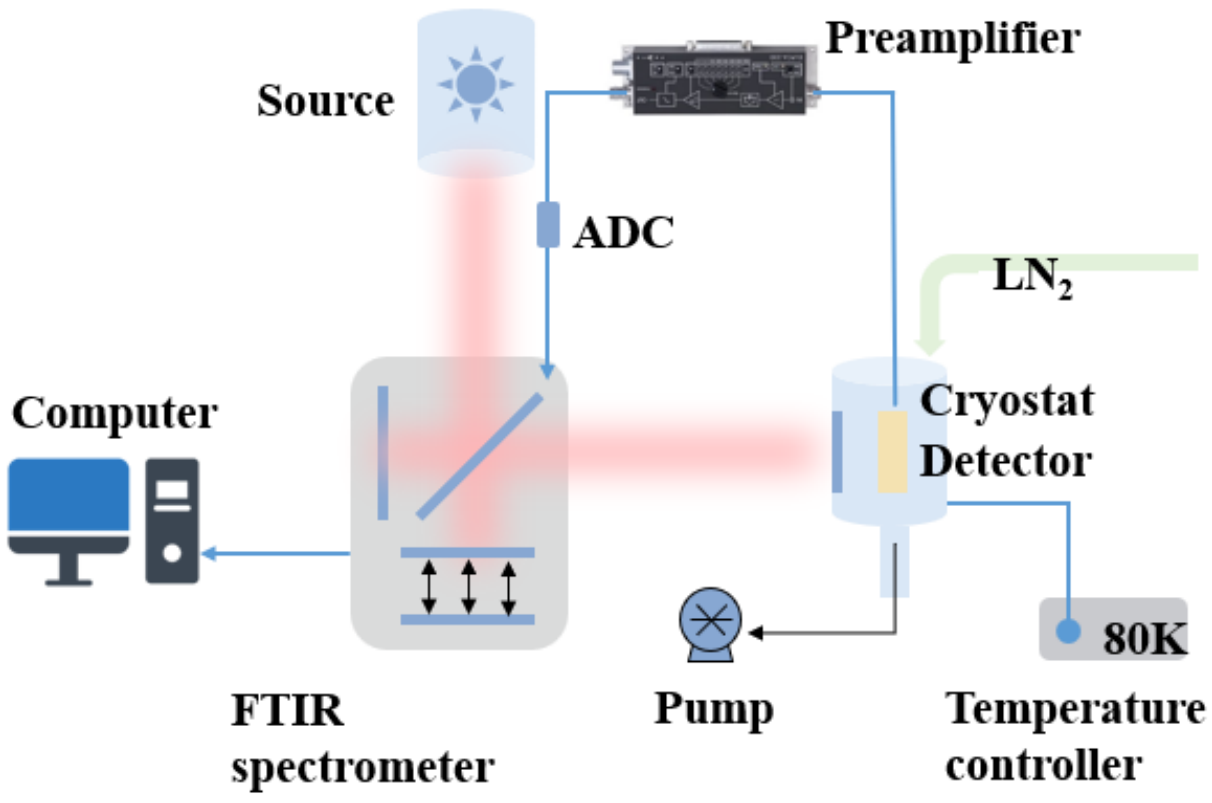

Fig. S6 Schematic of the measurement setup for spectral response

The spectral response was measured with an FTIR (Bruker 80v) as illustrated in Fig. S6 [51]. The detectors are packaged and mounted in the same dewar employed for the blackbody response measurement. To eliminate the influence of air molecules, all paths traversed by light were hermetically sealed and evacuated. Radiation emitted from the tungsten light was directed onto the detectors. Initial photocurrent signals were amplified by the amplifier SR560. Subsequently, the signal is translated into a digital signal by an AD converter, and then read out by the computer. This process yielded the original photocurrent $I_{\lambda 0}$ of the detectors. To enhance the signal-noise ratio, two different beam splitters were used. A $CaF_2$ beam splitter was used to measure the spectrum from 1 to 2.7μm while a KBr splitter was employed for the spectrum above 2.7μm. The background spectra $P_b$ of both wavebands were also acquired using the respective beam splitters and a HgCdTe detector. Subsequently, the spectral response of the device $R_\lambda$ could be calculated by the formula $R_\lambda = I_{\lambda 0} / P_b$. The spectral detectivity is calculated through $D_\lambda^* = D^* * g * R_\lambda / R_{\lambda_p}$, where $g$ factor represents the ratio between the peak responsivity/detectivity and the blackbody responsivity/detectivity. It can be calculated by the formula $g = \frac{R_{bb}}{R_{\lambda_p}} = \frac{D^*}{D_{\lambda_p}^*} = \frac{\int_0^\infty r(\lambda) R_\lambda d\lambda}{\int_0^\infty r(\lambda) d\lambda}$, in which $R_{\lambda_p}$, $D_{\lambda_p}^*$, $r(\lambda)$ represents peak responsivity, peak detectivity and blackbody emissivity, respectively.

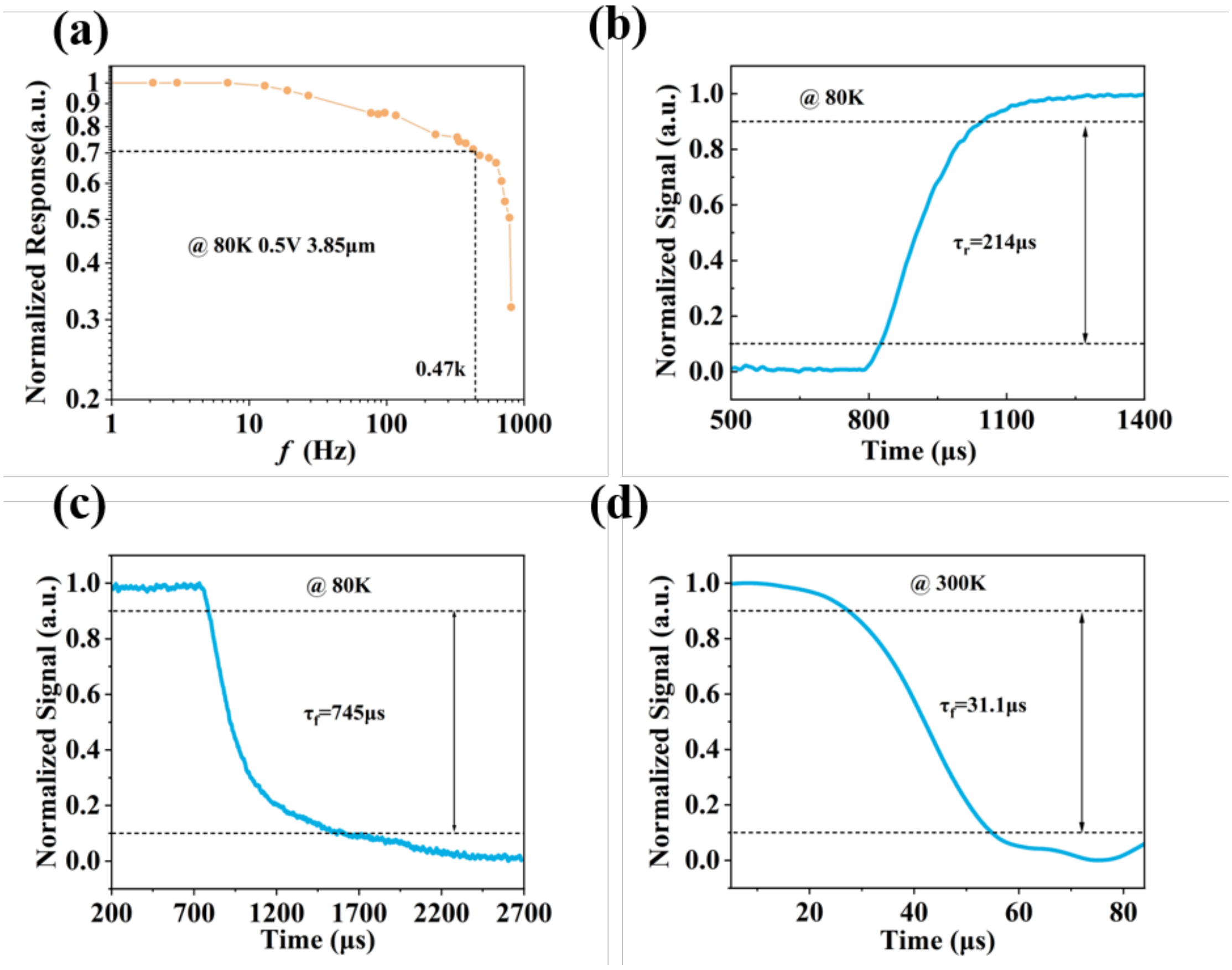

Fig. S7 (a) 3dB bandwidth of Si:S barrier detector measured at 1.4V under 80K (b) The rise time of Si:S barrier detector at 1.4V under 80K (c) The fall time of Si:S barrier detector at 1.4V under 80K (d) The fall time of Si:S barrier detector at 1.4V under 300K

The photoresponse time and the bandwidth were both measured with the same modulated laser. However, the method of photocurrent collection differed between the two measurements: an oscillator was utilized for response time measurement, whereas a lock-in amplifier was employed for bandwidth measurement. The fall time, defined as the duration between the 90% and 10% maximum values, serves as one of the metrics for response time in our case. Alternatively, the response time can be derived from its bandwidth. From the bandwidth analysis, the cut-off frequency $f_c$ is defined as the frequency at which the responsivity decreases to 0.707 times the maximum value. The relationship between cut-off frequency $f_c$ and response time is $\tau \cong \frac{0.35}{f_c}$[52]. The fall time observed in Fig. S7(b) aligns with the result obtained from Fig. S7(a), while the fall time depicted in Fig. S7(d) concurs with the outcome derived from Fig. 3(e).

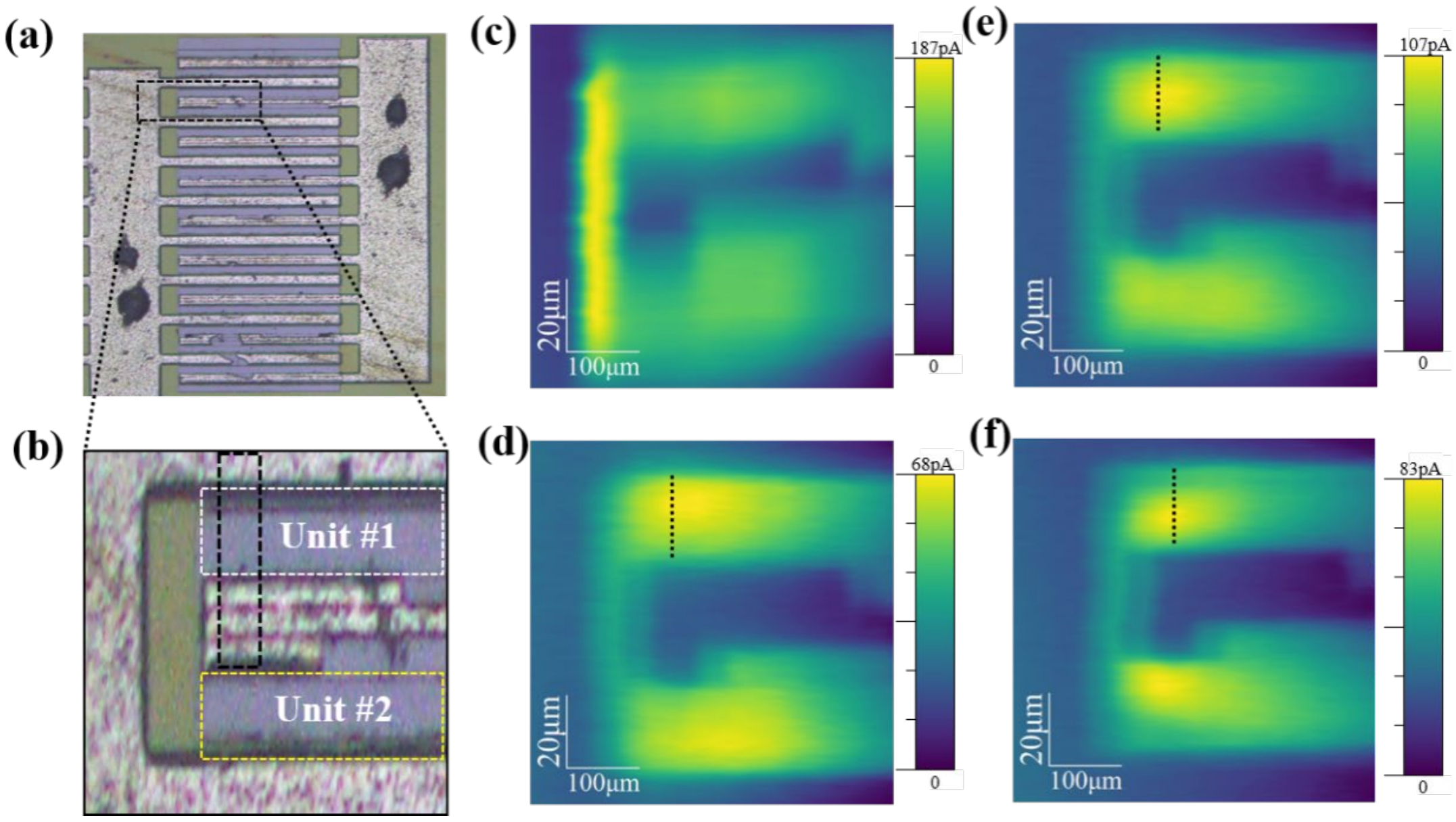

Fig. S8 (a)(b) The photo of the photocurrent mapping device (c) LBIC imaging of Si:S barrier detector under 637nm laser illumination at 4V (d-f) LBIC imaging of Si:S barrier detector under 1550nm laser illumination at 2V, 4V and 6V, respectively

The photograph depicting the detector for photocurrent mapping is presented in Fig. S8(a) and (b). In Fig. S8(b) to (f), the x and y coordinates are scaled at 5:1, contrasting with the 1:1 scale in Fig. S8(a). The outline of the detector area is not clear in Fig. S8(c) because the silicon beneath the electrode will also response to the 637nm light. The clarity of the outline improves in Fig. S8(d)~(f) since the interference from the area beyond the active region is reduced. Notably, the maximum photocurrent in Fig. S8(e) is 107 pA, which is the same magnitude as the maximum photocurrent in Fig. S8(c). This demonstrates the outstanding performance of our device in the infrared region.

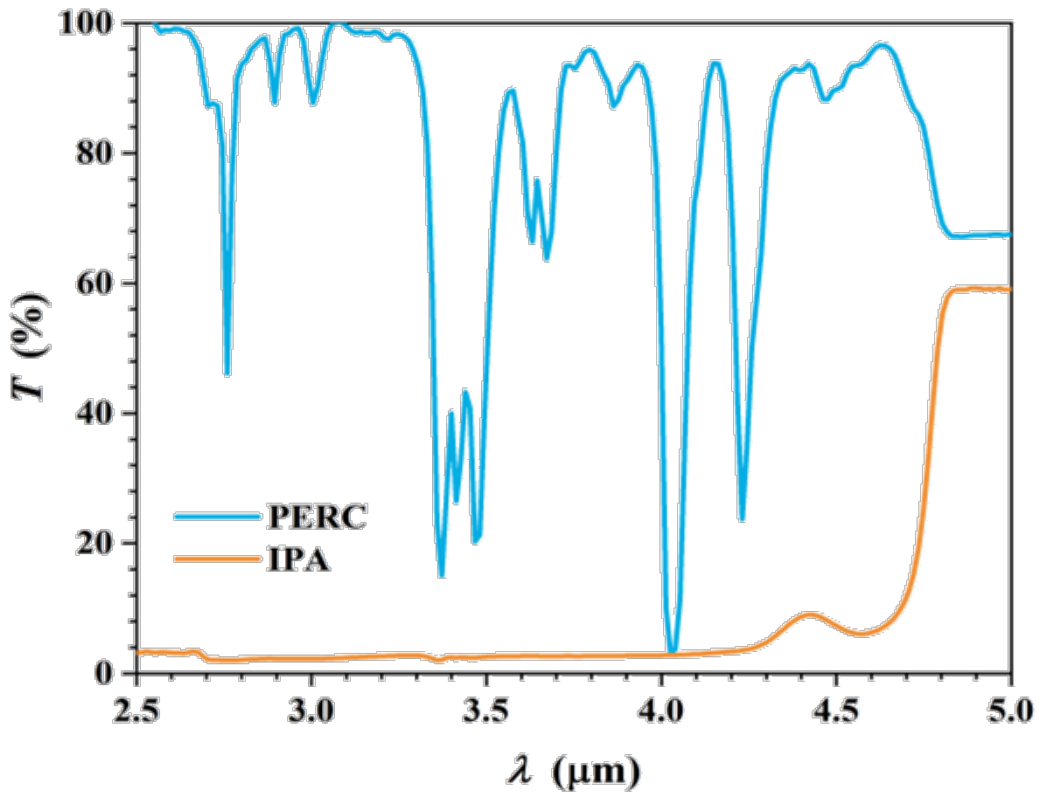

Fig. S9 The transmission spectrum of PERC and IPA measured by FTIR spectrometer

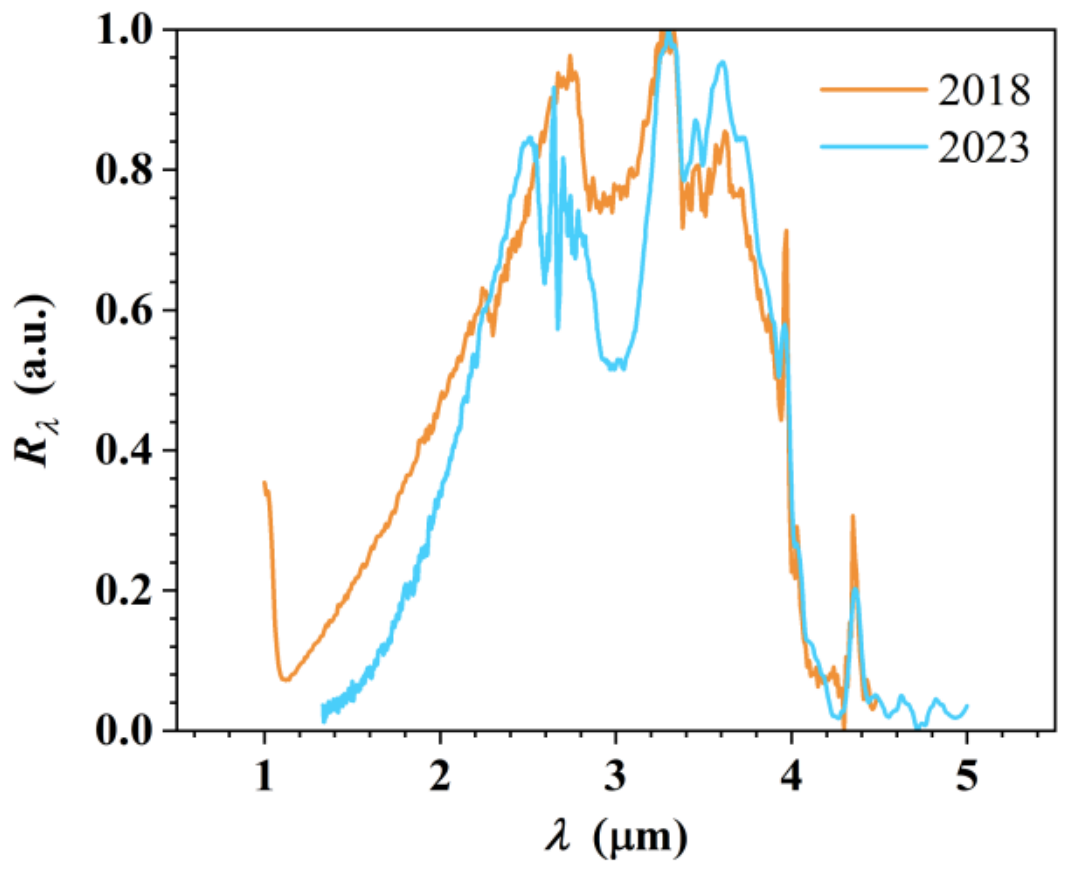

Fig. S10 The spectral response measured in 2018 and 2023, respectively.

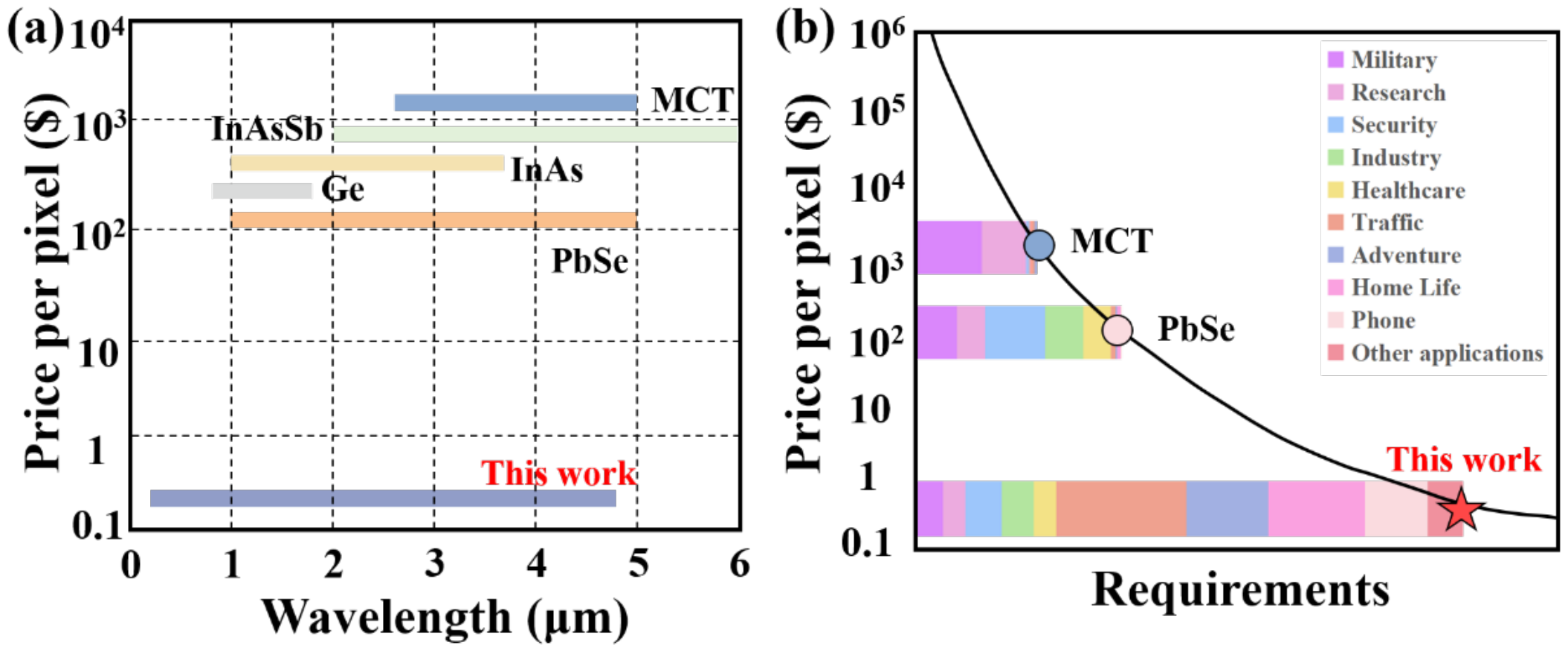

Fig. S11 (a) The fabrication cost of unit pixels with different material (b) Relationship between price and requirement of mid-infrared detectors

The fabrication cost is concluded from the data provided by Hamamatsu Photonics Corporation [53]. The relationship between price and requirement adheres to economic principles. Significant reductions in price by orders of magnitude can lead to the exploration of numerous unforeseen applications. Therefore, the practical application is much more than those given in Fig. S11(b).

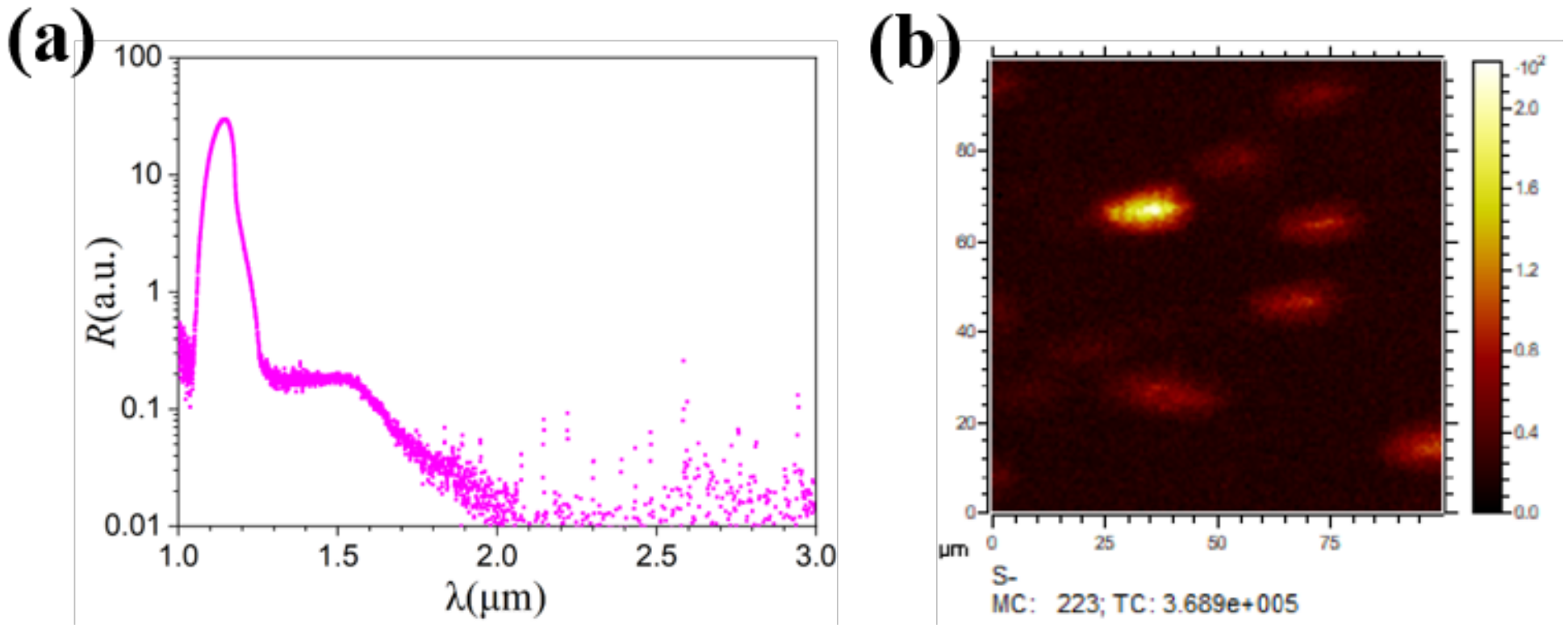

Fig. S12 (a)Spectral response of the failed samples (b)Top-view of sulfur distribution of the failed Si:S barrier infrared detectors

Table I Information about the reported mid-infrared detectors

| Material | Response Range(μm) | $D^*$ (cm・Hz$^{1/2}$/W) | $D_\lambda^*$ (cm・Hz$^{1/2}$/W) | $T_{Operating}$ (K) | Growth Technology | Maximum Size | Integration method |
|---|---|---|---|---|---|---|---|
| HgCdTe Photoconductor(VIGO)[54] | 1 to 5.3 | $1.0\times10^{10}$ | $2.0\times10^{10}$ | 293 | MBE | 5×5 cm (on ZnCdTe) | flip-chip |
| HgCdTe Photovoltaic(VIGO)[55] | 2.3 to 4.2 | $3.0\times10^{10}$ | $4.0\times10^{10}$ | 293 | MBE | 5×5 cm (on ZnCdTe) | flip-chip |
| InSb (HAMAMATSU) [56] | 1.6 to 5.5 | $3.0\times10^{10}$ | $1.6\times10^{11}$ | 77 | MBE | Ø10 cm (on InSb) | flip-chip |
| InAs (HAMAMATSU)[57] | 1 to 3.1 | - | $6.0\times10^{11}$ | 77 | MBE/MOCVD | Ø7.6 cm (on InAs) | flip-chip |
| InAsSb (HAMAMATSU)[58] | 2.3 to 4.9 | - | $1.0\times10^{9}$ | 293 | MBE/MOCVD | Ø10 cm (on GaSb) | flip-chip |
| nBn[59] | 2.5 to 6.0 | $1.5\times10^{10}$ | $3.65\times10^{10}$ | 160 | MBE | Ø7.6 cm (on Si) | flip-chip |
| T2SL (HAMAMATSU)[60] | 1 to 14.5 | - | $1.6\times10^{10}$ | 77 | MBE/MOCVD | Ø10 cm (on GaSb) | flip-chip |
| PbSe (N.E.P)[62] | 1 to 5.3 | - | $1.2\times10^{10}$ | 298 | Bridgman-Stockbarger | 10×10 cm (on PbSe) | deposition |
| HgTe QD[63] | 1.4 to 3.7 | $1.0\times10^{9}$ | - | 300 | Liquid-Phase Peptide Synthesis | - | spin coating |
| HgSe QD[64] | 3.0 to 5.0 | - | $1.7\times10^{9}$ | 80 | Liquid-Phase Peptide Synthesis | - | spin coating |
| bP[65] | 1.5 to 4.0 | $7.0\times10^{10}$ | $1.1\times10^{10}$ | 300 | CVD | 1×1 cm (on Mica) | transfer |
| bAsP[66] | 2.0 to 4.3 | $>4.9\times10^{9}$ | $9.2\times10^{9}$ | 300 | Micromechanical stripping method | 25×25 μm (on Si) | transfer |
| Te[67] | 0.64 to 3.8 | $1.11\times10^{8}$ | $3.69\times10^{8}$ | 300 | CVD | 30×50 μm (on $SiO_2$) | transfer |
| $PdSe_2$[68] | 0.4 to 4.05 | | $1.3\times10^{9}$ @1.06μm | 300 | Micromechanical stripping method | 50×50 μm (on $SiO_2$/Si) | transfer |
| This work | 1.12 to 4.4 | $6.25\times10^{9}$ | $1.4\times10^{11}$ | 90 | Czochralski | Ø30 | monolithic |
| | | $1.93\times10^{8}$ | $4.4\times10^{9}$ | 210 | | | |

Table II Information about the reported silicon-based infrared detectors

| Material | Dopant Concentration($cm^{-3}$) | Detector Type | $T_{Operating}$ (K) | Response Range(μm) | $\lambda_p$(μm) | Responsivity(A /W) |
|---|---|---|---|---|---|---|
| Si:Au[69] | $5\times10^{20}$ | Photoconductive | 300 | 2.2 | - | - |
| Si:Zn[70] | $10^{19}$ | Photoconductive | - | 1.1-2.5 | - | $6.8\times10^{-4}$ @1550nm |
| Si:Ti[71] | $10^{20}$ | Photovoltaic | - | 0.4-1.55 | 0.95 | $3.42\times10^{-3}$ @1550nm |
| Si:Se[72] | $9\times10^{20}$ | Photoconductive | 300 | 1.44~3.1 | - | $72\pm3\times10^{-6}$ @1550nm |
| Si:Te[73] | $1.2\times10^{20}$ | Photoconductive | 300 | 0.6-1.1 | 1.12 | $56.8\times10^{-3}$ @1550nm |
| Si:Er[74] | $10^{21}$ | Photoconductive | 300 | 1.25-1.55 | 1.31 | $1.65\times10^{-4}$ @1310nm |
| Si:S[75] | $10^{16}$ | Photoconductive | - | 3.36-3.74 | - | $2.2\times10^{-3}$ @3740nm |
| This work | $2\times10^{17}$ | Photoconductive | 210 | 1.12-4.4 | 3.3 | 2.05@3300nm |
| | | | 300 | 1.12-4.4 | | $4.4\times10^{-2}$ @3133nm |